\documentclass[
 reprint,
 amsmath,amssymb,
 aps,nofootinbib
]{revtex4-2}
\usepackage[T1]{fontenc}
\usepackage{graphicx}
\usepackage{dcolumn}
\usepackage{bm}
\usepackage{subfig}
\usepackage{color}
\usepackage{multirow}
\begin{document}

\preprint{APS/123-QED}

\title{Double strangeness pentaquarks and other exotic hadrons in the $\Xi_b \to \Xi\ J/\Psi\ \phi$ decay}

\author{J.A. Mars\'e-Valera}
\author{V.K. Magas}
\author{A. Ramos}
\affiliation{
$ ^1$ Departament de Física Quàntica i Astrofísica (FQA), Universitat de Barcelona (UB),  c. Martí i Franqués, 1, 08028 Barcelona, Spain\\
$ ^2$ Institut de Ciències del Cosmos (ICCUB), Universitat de Barcelona (UB), c. Martí i Franqués, 1, 08028 Barcelona, Spain}
\date{\today}

\begin{abstract}
We study the possibility that four $\Xi$ resonances $(\Xi(1620)$, $\Xi(1690)$, $\Xi(1820)$, $\Xi(1950))$ could correspond to pentaquark states, in the form of meson-baryon bound systems. We also explore the possible existence of doubly strange pentaquarks with hidden charm $(P_{css})$ and find two candidates structured in a similar form, at energies of $4493~{\rm MeV}$ and $4630~{\rm MeV}$. The meson-baryon interaction is built from t-channel meson exchange processes which are evaluated using effective Lagrangians. Moreover we analyse the $\Xi_b\to\Xi~J/\psi~\phi$ decay process, which permits exploring the existence of the heavy double strange pentaquark, as well as other exotic hadrons, in the three different two-body invariant mass spectra of the emitted particles. In the $J/\psi\phi$ mass spectrum, we analyse the nature of the $X(4140)$ and $X(4160)$ resonances. In the $J/\psi\Xi$ invariant mass spectrum, we study the signal produced by the double strange pentaquark, where we conclude that it has a good chance to be detected in this reaction if its mass is around $4580-4680~{\rm MeV}$. Finally, in the $\phi\Xi$ spectrum we study the likelihood to detect the $\Xi(2500)$ state.      
\end{abstract}


\maketitle

\section{Introduction}
\label{sec:intro}

In the last few decades lots of different multi-quark states have been observed experimentally, showing the existence of more complex structures than the conventional mesons and baryons, made of a quark-antiquark pair and three quarks respectively. This fact motivated many theoretical groups to study models that can generate hadrons beyond the standard structures, which increased our understanding of the hadron interaction. One kind of these exotic configurations proposes baryons composed by five quarks, but structured in a quasi-bound state of an interacting meson-baryon pair.

One example of the success of this type of models is provided by the $\Lambda(1405)$ resonance, for which the quark models systematically over-predicted its mass. Instead, dedicated studies of the meson-baryon interaction obtained from chiral effective Lagrangians predict the $\Lambda(1405)$ to be a $\Bar{K}N$ molecule \cite{Dalitz:1960du,Kaiser:1995eg,Oset:1997it}, what actually allows to interpret the $\Lambda(1405)$ as the first observed pentaquark.
Furthermore, the chiral models with unitarization in coupled channels have predicted its double-pole nature~\cite{Oller:2000fj,Jido:2003cb}, which can be seen from comparing different experimental line shapes~\cite{Magas:2005vu}, and which is now commonly accepted in the field and appears in the PDG ~\cite{PDG}.
After the successful interpretation of the $\Lambda(1405)$, many groups tried to extend this kind of models to other  spin, isospin and flavour sectors, finding a more natural way to explain different states, like the $f_0(500)$ and the $a_0(980)$, which are described as meson-meson quasi-bound states \cite{Oller:1997ti,Oller:1998hw,Pelaez:2015qba}. 

More recent experimental results, like the discovery at LHCb \cite{LHCb:2015yax,LHCb:2019kea} of four excited resonances  of the nucleon ($P_c(4312)$, $P_c(4380)$, $P_c(4440)$ and $P_c(4457))$, seen in the invariant mass distribution of $J/\psi p$ pairs from the $\Lambda_b$ decay, and the more recent report of LHCb \cite{M.Wang:Psc} which, analysing the $\Xi_b$ decay, finds evidence of the existence of a pentaquark with strangeness $(P_{cs}(4459))$ on the invariant mass distribution of $J/\psi\Lambda$ pairs, clearly establish the need for including a $c\Bar{c}$ pair excitation in order to reproduce the high masses of these states.

Baryons with hidden charm, structured as a meson-baryon molecule, had been predicted in some earlier works \cite{Wu:2010vk,Hofmann:2005sw,Yuan:2012wz,Xiao:2013yca,Garcia-Recio:2013gaa}, prior to their discovery, but the existence of a hidden charm pentaquark with double strangeness was not addressed until recently. Adopting a one-boson-exchange model to derive effective interaction potentials among the hadrons, the work of  \cite{Wang:2020bjt} found a pair of slightly bound meson-baryon molecules with double strangeness, coupling strongly to $\Xi_c^\prime\Bar{D}_s^*$  and $\Xi_c^*\Bar{D}_s^*$ states respectively, albeit with the use of somewhat hard and unrealistic cut-off values of around 2 GeV.  In contrast, the works based on a t-channel vector-meson exchange interaction between mesons and baryons in the hidden charm sector did not find double strangeness pentaquark molecules \cite{Wu:2010vk}. However, a recent study pointed out that, in spite of the fact that these models predict weakly attractive or even repulsive meson-baryon interactions, a strong coupled-channel effect does indeed produce enough attraction to generate bound pentaquarks  \cite{Marse-Valera:2022khy,Magas:2024biu} with strangeness $S=-2$, $P_{css}$.
The prediction of such pentaquarks has been corroborated in Ref. \cite{Roca:2024nsi}, where a similar
t-channel vector-meson interaction was employed, but deriving the baryon-baryon-vector meson vertices from the spin and flavour wave functions rather than from SU(4) arguments as in \cite{Marse-Valera:2022khy}.
We note that  $P_{css}$ pentaquarks have also been predicted within quark-model approaches \cite{Anisovich:2015zqa,Ortega:2022uyu} or employing sum rule techniques  \cite{Azizi:2021pbh,Wang:2022neq}.

In the recent work of Ref. \cite{Oset:2024fbk} the authors have proposed the $\Xi_b^0\to \eta \eta_c \Xi^0$ and $\Omega_b^-\to K^- \eta_c \Xi^0$ decay processes to look for the $P_{css}$ with $I(J^P)=\frac{1}{2}(\frac{1}{2}^-)$ generated via the pseudoscalar-baryon interaction. The enhanced intensity of the $\Omega_b^-$ decay process establishes it as having a better chance for observation according to their study, but it is also much more difficult experimentally.

This paper has a twofold purpose. On the one hand, we present the details of the formalism that produces dynamically generated  $P_{css}$ states, which were omitted in a previously published short letter \cite{Marse-Valera:2022khy}, while showing at the same time the prediction of the model for the lighter energy range of the $S=-2$  sector (including some other resonances generation).
This sector has experienced a renewed interest due to recent data from the Belle experiment \cite{Belle:2018lws}, studied, for example, in \cite{Nishibuchi:2023acl, Feijoo:2023wua}, and even more recently from the femtoscopy analysis of the ALICE collaboration \cite{ALICE:2023wjz}, which are analyzed in particular in Ref. \cite{Sarti:2023wlg}.

On the other hand, we propose a decay process, $\Xi_b\to\Xi~J/\psi~\phi$, in which the predicted pentaquark can be seen in the $J/\psi\Xi$ spectrum.  As we will discuss in section \ref{dalitz}, this reaction is interesting because it allows to detect and study some other exotic hadrons too, namely a family of $X$ resonances from $X(4140)$ till $X(4700)$ and $\Xi(2500)$.

\section{Meson-Baryon interactions for S=-2 sector}\label{sec:MBS}
\subsection{Meson-Baryon interaction}

\begin{figure}[h!]
     \includegraphics[scale=0.25]{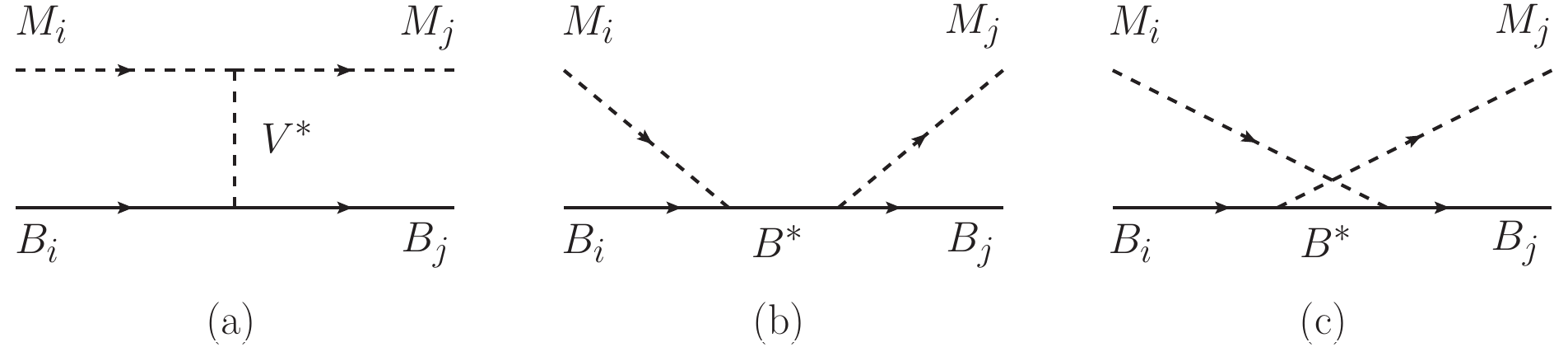}
     \caption{
     	Leading order tree-level diagrams contributing to the meson-baryon interaction:  t-channel term (a), s-channel term (b), and u-channel term (c).  Baryons and mesons are depicted by solid and dashed lines, respectively.}
     \label{Fig:treediagr}
 \end{figure}

The model of the meson-baryon interaction used in this work is based on the tree-level diagrams of Fig. \ref{Fig:treediagr}. We will only consider the s-wave amplitude, for which the most important contribution comes from the t-channel term (Fig. \ref{Fig:treediagr}(a)). The s- and u- channel terms (Fig. \ref{Fig:treediagr}(b) and (c)) contribute mostly to the p-wave amplitude and they may have effects at higher energies. In Ref. \cite{Oller:2000fj} the contribution from these terms in the $S=-1$ sector were calculated, and were found to be around $20\%$ of that of the dominant t-channel around $200~{\rm MeV}$ above the threshold. We can expect that in the $S=-2$ sector studied here, these terms will contribute even less, as the intermediate baryon is more massive, and therefore they will be neglected.

To describe the interaction we employ effective Lagrangians that describe the couplings of the vector meson to pseudoscalar mesons $(VPP)$ and baryons $(VBB)$, which are obtained using the hidden gauge formalism and assuming $SU(4)$ symmetry \cite{Hofmann:2005sw}:
\begin{equation}\label{eq:vertexVPP}
\mathcal{L}_{VPP}=ig\langle\left[\partial_\mu\phi, \phi\right] V^\mu\rangle,
\end{equation}
\begin{equation}\label{eq:vertexBBV}
\mathcal{L}_{VBB}=\frac{g}{2}\sum_{i,j,k,l=1}^4\bar{B}_{ijk}\gamma^\mu\left(V_{\mu,l}^{k}B^{ijl}+2V_{\mu,l}^{j}B^{ilk}\right),
\end{equation}
where $\phi$ and $V_\mu$ represent the 16-plet pseudo-scalar field and the 16-plet vector field, respectively, and $\langle~\rangle$ denotes the $SU(4)$ trace in flavour space. The $g$ factor is a coupling constant which is related to the pion decay constant $(f=93~{\rm MeV})$ and a representative mass of a light vector meson from nonet $(m_V)$ through the following relation,
\begin{equation}\label{eq:g}
    g=\frac{m_V}{2f}.
\end{equation}

Using the $VPP$ and $VBB$ vertices in the t-exchange diagram of Fig. \ref{Fig:treediagr} (a) one can obtain the s-wave interaction kernel $V_{ij}$ \cite{Hofmann:2005sw} for the pseudoscalar meson baryon $(PB)$ interaction:
\begin{eqnarray}\label{Vij-1}
    V_{ij}=g^2\sum_vC^v_{ij}\bar{u}(p_j)\gamma^\mu u(p_i)\frac{1}{t-m_v^2}\times \nonumber \\
    \Bigg[(k_i+k_j)_\mu-\frac{k_i^2-k_j^2}{m_v}(k_i+k_j)_\mu\Bigg],
\end{eqnarray}
where $p_{i/j}$ and $k_{i/j}$ denote the four-momentum of the baryons and mesons, respectively, and $i$, $j$ are the incoming and outgoing meson-baryon channels. The mass of the vector meson interchanged is $m_v$, and we approximate it as $m_v=m_V$, being equal for all of the vector mesons without charm quark and/or antiquark. For the charmed mesons we add a factor $\kappa_c=(m_V/m_V^c)^2=1/4$ for $D^*$, $D_s^*$ mesons and $\kappa_{cc}=1/9$ for the $J/\psi$ in order to take account of their higher mass. This simplifies Eq. 
(\ref{Vij-1}) into
\begin{equation}\label{Vij-2}
    V_{ij}=-C_{ij}\frac{1}{4f^2}\bar{u}(p_j)\gamma^\mu u(p_i)(k_i+k_j)_\mu,
\end{equation}
where we have taken the limit $t\ll m_V$, which reduces the t-channel diagram to a contact term, and we have included the $\kappa_c$ and $\kappa_{cc}$ factors in the $C_{ij}$ matrix of coefficients, which will be provided later. This expression can be further simplified using the Dirac algebra up to $\mathcal{O}(p^2/M^2)$, what leads to:
\begin{equation}\label{eq:Vij}
 V_{ij}(\sqrt{s})=-C_{ij}\frac{1}{4f^2}\left(2\sqrt{s}-M_i-M_j\right) N_i N_j,
\end{equation}
where $M_i$, $M_j$ are the masses of the baryons in the channels $i$ and $j$, $N_i$ and $N_j$ are the normalization factors $N_i=\sqrt{(E_i+M_i)/2M_i}$ and $E_i$, $E_j$ are the energies of the corresponding baryons. We note that although $SU(4)$ symmetry has been employed to obtain the $C_{ij}$ coefficients, our interaction potential is not $SU(4)$ symmetric, since we use the physical masses of mesons and baryons.

The interaction of vector mesons with baryons is built in a similar way and involves a three-vector $VVV$ vertex obtained from
\begin{equation}\label{eq:vertexVVV}
\mathcal{L}_{VVV}=ig\langle[V^\mu,\partial_\nu V_\mu]V^\nu\rangle.
\end{equation}

One can see that in this case we can arrive to a similar expression as Eq.~(\ref{eq:Vij}), but multiplied by  the product of the  polarization vectors, $\Vec{\epsilon}_i\cdot\Vec{\epsilon}_j$:
\begin{equation}\label{eq:VijVB}
 V_{ij}(\sqrt{s})=-C_{ij}\Vec{\epsilon}_i\cdot\Vec{\epsilon}_j\frac{1}{4f^2}\left(2\sqrt{s}-M_i-M_j\right) N_i N_j.
\end{equation}

\subsection{The $I=1/2~S=-2$ sector}

In the $I=1/2~S=-2$ sector we have $9$ possible pseudoscalar-meson baryon (PB) channels, namely $\pi\Xi(1456)$, $\bar{K}\Lambda(1611)$, $\bar{K}\Sigma(1689)$, $\eta\Xi(1866)$, $\eta'\Xi(2276)$, $\eta_c\Xi(4302)$, $\bar{D}_s\Xi_c(4437)$, $\bar{D}_s\Xi_c'(4545)$, $\bar{D}\Omega_c(4565)$, where the values in parenthesis denote their thresholds. As can be seen, the channels with hidden charm are about $2~{\rm GeV}$ more massive compared with the lighter channels. Therefore we can expect these two regions to be essentially independent, making only a rather small effect on each other. The values of the $C_{ij}$ coefficients for the nine channels are displayed in Table \ref{Tab:CijPB}. 

\begin{table}[h!]
  \begin{center}
    \begin{tabular}{c|ccccc|cccc}
        \hline\\ [-0.35cm]
         & $\pi\Xi$ & $\bar{K}\Lambda$ & $\bar{K}\Sigma$ & $\eta\Xi$ & $\eta'\Xi$ & $\eta_c\Xi$ & $\bar{D}_s\Xi_c$ & $\bar{D}_s\Xi_c'$ & $\bar{D}\Omega_c$\\
        \hline
        $\pi\Xi$ & $2$ & $\frac{3}{2}$ & $\frac{1}{2}$ & $0$ & $0$ & $0$ & $0$ & $0$ & $\sqrt{\frac{3}{2}}\kappa_c$\\
        $\bar{K}\Lambda$ & & $0$ & $0$   & $-\frac{3}{2}$ & $0$ & $0$ & $-\frac{1}{2}\kappa_c$ & $-\frac{\sqrt{3}}{2}\kappa_c$ & $0$\\
        $\bar{K}\Sigma$ & & & $2$ & $\frac{3}{2}$ & $0$ & $0$ & $\frac{3}{2}\kappa_c$ & $-\frac{\sqrt{3}}{2}\kappa_c$ & $0$\\
        $\eta\Xi$& & & & $0$ & $0$ & $0$ & $\kappa_c$ & $\frac{1}{\sqrt{3}}\kappa_c$ & $\frac{1}{\sqrt{6}}\kappa_c$\\
        $\eta'\Xi$ & & & & & $0$ & $0$ & $-\frac{1}{\sqrt{2}}\kappa_c$ & $-\frac{1}{\sqrt{6}}\kappa_c$ & $\frac{1}{\sqrt{3}}\kappa_c$\\
        \hline
        $\eta_c\Xi$ & & & & & & $0$ & $\sqrt{\frac{3}{2}}\kappa_c$ & $\frac{1}{\sqrt{2}}\kappa_c$ & $-\kappa_c$\\
        $\bar{D}_s\Xi_c$ & & & & & & & $\kappa_{cc}-1$ & $0$ & $0$\\
        $\bar{D}_s\Xi_c'$ & & & & & & & & $\kappa_{cc}-1$ & $-\sqrt{2}$\\
        $\bar{D}\Omega_c$ & & & & & & & & & $\kappa_{cc}$ \\
    \end{tabular}
  \end{center}
\caption{Coefficients $C_{ij}$ of the $PB$ interaction in the $I=1/2$, $S=-2$ sector}    
\label{Tab:CijPB}
\end{table}

In the vector meson baryon (VB) sector, the allowed channels are $\rho\Xi(2089)$, $\bar{K}^*\Lambda(2010)$, $\bar{K}^*\Sigma(2087)$, $\omega\Xi(2101)$, $\phi\Xi(2338)$, $J/\psi\Xi(4415)$, $\bar{D}^*_s\Xi_c(4581)$, $\bar{D}^*_s\Xi_c'(4689)$, $\bar{D}^*\Omega_c(4706)$, which can again be separated in two groups, one with the light channels and the other with the heavy ones. In this case we can obtain the $C_{ij}$ coefficients directly from those in Table \ref{Tab:CijPB} using the following correspondences,
\begin{eqnarray}
  && \pi \rightarrow \rho, ~~ K\rightarrow K^\ast, ~~ \bar{K}\rightarrow\bar{K}^\ast, ~~ D\rightarrow D^\ast,  ~~\bar{D}\rightarrow\bar{D}^\ast \nonumber \\
  && \eta_c\rightarrow J/\psi,  \frac{1}{\sqrt{3}}\eta+\sqrt{\frac{2}{3}}\eta^\prime\rightarrow\omega,  -\sqrt{\frac{2}{3}}\eta+\frac{1}{\sqrt{3}}\eta^\prime\rightarrow\phi\ ,
 \label{eq:eta_phi}
\end{eqnarray}
and these new coefficients are listed in Table \ref{Tab:CijVB}.\\

\begin{table}[h!]
  \begin{center}
    \begin{tabular}{c|ccccc|cccc}

        \hline\\ [-0.35cm]
         & $\rho\Xi$ & $\bar{K}^*\Lambda$ & $\bar{K}^*\Sigma$ & $\omega\Xi$ & $\phi\Xi$ & $J/\psi\Xi$ & $\bar{D}^*_s\Xi_c$ & $\bar{D}^*_s\Xi_c'$ & $\bar{D}^*\Omega_c$\\
        \hline
        $\rho\Xi$ & $2$ & $\frac{3}{2}$ & $\frac{1}{2}$ & $0$ & $0$ & $0$ & $0$ & $0$ & $\sqrt{\frac{3}{2}}\kappa_c$\\
        $\bar{K}^*\Lambda$ & & $0$ & $0$   & $-\frac{\sqrt{3}}{2}$ & $\sqrt{\frac{3}{2}}$ & $0$ & $-\frac{1}{2}\kappa_c$ & $-\frac{\sqrt{3}}{2}\kappa_c$ & $0$\\
        $\bar{K}^*\Sigma$ & & & $2$ & $\frac{\sqrt{3}}{2}$ & $-\sqrt{\frac{3}{2}}$ & $0$ & $\frac{3}{2}\kappa_c$ & $-\frac{\sqrt{3}}{2}\kappa_c$ & $0$\\
        $\omega\Xi$& & & & $0$ & $0$ & $0$ & $0$ & $0$ & $\frac{1}{\sqrt{2}}\kappa_c$\\
        $\phi\Xi$ & & & & & $0$ & $0$ & $-\sqrt{\frac{3}{2}}\kappa_c$ & $-\frac{1}{\sqrt{2}}\kappa_c$ & $0$\\
        \hline
        $J/\psi\Xi$ & & & & & & $0$ & $\sqrt{\frac{3}{2}}\kappa_c$ & $\frac{1}{\sqrt{2}}\kappa_c$ & $-\kappa_c$\\
        $\bar{D}_s\Xi_c$ & & & & & & & $\kappa_{cc}-1$ & $0$ & $0$\\
        $\bar{D}_s\Xi_c'$ & & & & & & & & $\kappa_{cc}-1$ & $-\sqrt{2}$\\
        $\bar{D}\Omega_c$ & & & & & & & & & $\kappa_{cc}$ \\

    \end{tabular}
  \end{center}
    \caption{Coefficients $C_{ij}$ of the $VB$ interaction  in the $I=1/2$, $S=-2$ sector}   
    \label{Tab:CijVB}
\end{table}

Note that in the light $PB$ sector there are formally only $4$ coupled channels, due to the fact that all $C_{5j}$ coefficients with $j\leq 5$ are zero, while this is not the case for the light $VB$ sector, which remains with $5$ coupled channels.

\subsection{Dynamically generated resonances}

\begin{figure}[h!]
    \includegraphics[scale=0.33]{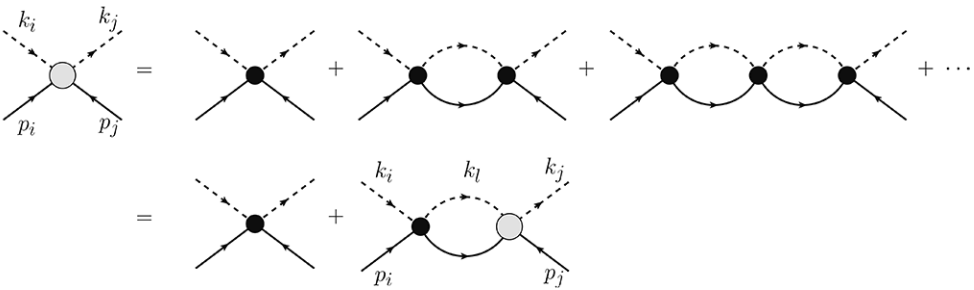}
    \caption{Diagrams that represent the Bethe-Salpeter equation. The empty circle corresponds to the $T_{ij}$ matrix elements, the black ones represent the $V_{ij}$ potential and the loops are the $G_l$ propagator functions.}
    \label{fig:BetSalfig}
\end{figure}

The sought resonances are dynamically generated as poles of the scattering amplitude $T_{ij}$, which is unitarized via the on-shell Bethe-Salpeter equation in coupled channels. This corresponds to performing the resummation of loop diagrams to an infinite order (see Fig. \ref{fig:BetSalfig}),
\begin{equation}\label{eq:Tij1}
T_{ij}=V_{ij}+V_{il}G_lV_{lj}+V_{il}G_lV_{lk}G_kV_{kj}+...=V_{ij}+V_{il}G_lT_{lj}
\end{equation}
where $G_l$ is the loop function, which is given by
\begin{equation}\label{eq:Gl}
G_l=i\int\frac{d^4q}{(2\pi)^4}\frac{2M_l}{(P-q)^2-M_l^2+i\epsilon}\frac{1}{q^2-m_l^2+i\epsilon}.
\end{equation}
The masses $M_l$ and $m_l$ correspond to those of the baryon and meson of the $l$ channel, respectively, $P=p+k=(\sqrt{s},0)$ is the total four momenta in the c.m. frame, and $q$ denotes the four momentum in the intermediate loop. We factorize the $V$ and $T$ matrices on-shell out of the integrals in Eq.~(\ref{eq:Tij1}) \cite{Montana:2017kjw}, hence obtaining the following algebraic matrix expression for the scattering amplitude,
\begin{equation}\label{eq:Tij2}
    T=(1-VG)^{-1}V.
\end{equation}

Note that, in searching the poles of the VB amplitudes, one can consider only the transverse mode of the vector polarization vector, as discussed in Appendix B of Ref.~\cite{Roca:2005nm}. Moreover, disregarding the small contribution of the $q_i q_j/M_V^2$ term of the virtual vector meson propagator, one can factorize the product $\Vec{\epsilon}_i\Vec{\epsilon}_j$ out from all terms in the Bethe-Salpeter equation.

\subsection{Loop function}

It is important to know that the loop function in Eq.~(\ref{eq:Gl}) diverges logarithmically, so we must renormalize it. We  can employ the \textit{cut-off} method, which consists in changing the infinity at the upper $q$ limit to one large enough cut-off momentum $\Lambda$
\begin{equation}\label{eq:Glcut}
 G_l^{cut}=\int_0^\Lambda \frac{q^2dq}{4\pi^2}\frac{2M_l(E_l+\omega_l)}{E_l\omega_l[(P^0)^2-(E_l+\omega_l)^2+i\epsilon]}, 
\end{equation}
 where $E_l=\sqrt{\vec{q}^{\,2}+M_l^2}$ and $\omega_l=\sqrt{\vec{q}^{\,2}+m_l^2}$. 

A different option is using the \textit{dimensional regularisation} scheme, which gives rise to the following analytic expression
\begin{equation}\label{eq:GmatrixDR}
\begin{aligned}
 G_{l}=&\frac{2M_l}{16\pi^2}\Big\{ a_l(\mu)+\ln\frac{M_l^2}{\mu^2}+\frac{m_l^2-M_l^2+s}{2s}\ln\frac{m_l^2}{M_l^2}+ \\ 
  &+\frac{q_l}{\sqrt{s}}\left[\ln\left(s-(M_l^2-m_l^2\right)+2q_l\sqrt{s})\right. \\
  &\phantom{\frac{q_l}{\sqrt{s}}~~}+\ln\left(s+(M_l^2-m_l^2\right)+2q_l\sqrt{s})\\
  &\phantom{\frac{q_l}{\sqrt{s}}~~}-\ln\left(-s+(M_l^2-m_l^2\right)+2q_l\sqrt{s})\\
  &\phantom{\frac{q_l}{\sqrt{s}}~~}\left.-\ln\left(-s-(M_l^2-m_l^2\right)+2q_l\sqrt{s}) \right] \Big\},
\end{aligned}
\end{equation}
were $q_l$ is the on-shell three-momentum of the meson in the loop, and we have introduced a subtraction constant, $a(\mu)$, at the regularisation scale $\mu$. In this work we will take $\mu=1000~{\rm MeV}$ consistently with previous approaches \cite{Montana:2017kjw}. Although these two expression for the loop function, (\ref{eq:Glcut}) and (\ref{eq:GmatrixDR}), give rather different results at low (sub-threshold) and high energies, they can be rather similar in the threshold region of the corresponding channels. The value of the subtraction constant that produces the same loop function at threshold as that obtained with a cut-off $\Lambda$ is determined from
\begin{equation}\label{eq:a(mu)}
 a_l({\mu})= \frac{16\pi^2}{2M_l}\left(G_{l}^\text{cut}(\Lambda)-G_{l}(\mu,a_l=0)\right).
\end{equation}

Some mesons in the considered channels have a large width $(\Gamma_\rho=149.4~{\rm MeV}$ and $\Gamma_{K^*}=50.5~{\rm MeV}$), and this is implemented by convoluting $G_l$ in Eq.~(\ref{eq:GmatrixDR}) with the mass distribution of the particle. This method has been used in \cite{Montana:2017kjw,Oset:2010tof} and the resulting loop function for the channels involving wide mesons becomes: 
\begin{eqnarray}
 \tilde{G}_{l}(s)=-\frac{1}{N}\int_{(m_l-2\Gamma_l)^2}^{(m_l+2\Gamma_l)^2} & \frac{d\tilde{m}_l^2}{\pi}{\rm\, Im\,}\frac{1}{\tilde{m}_l^2-m_l^2+i\,m_l\Gamma(\tilde{m}_l)} \nonumber \\
 \times G_{l}\left(s,\tilde{m}_l^2,M_l^2\right),
\label{eq:loop_conv}
\end{eqnarray}
where  we extend the integration limits up to twice the meson width on either side of the mass, and the normalisation factor $N$ is
\begin{equation}
 N=\int_{(m_l-2\Gamma_l)^2}^{(m_l+2\Gamma_l)^2}d\tilde{m}_l^2\left(-\frac{1}{\pi}\right){\rm\, Im\,}\frac{1}{\tilde{m}_l^2-m_l^2+i\,m_l\Gamma(\tilde{m}_l)} \ .
\end{equation}
The energy-dependent width is given by
\begin{equation}\label{eq:G_kallen}
 \Gamma(\tilde{m}_l)=\Gamma_l\frac{m_l^5}{\tilde{m}_l^5}\frac{\lambda^{3/2}(\tilde{m}_l^2,m_1^2,m_2^2)}{\lambda^{3/2}(m_l^2,m_1^2,m_2^2)}\,\theta(\tilde{m}_l-m_1-m_2),
\end{equation}
where $m_1$ and $m_2$ are the masses of the lighter mesons into which the vector meson decays and $\lambda$ is the K\"{a}ll\'{e}n function $\lambda(x,y,z)=(x-(\sqrt{y}+\sqrt{z})^2)(x-(\sqrt{y}-\sqrt{z})^2)$.

Then, the obtained scattering amplitude ($T$) can be analytically continued to the complex plane of $s$. The dynamically generated resonances appear as poles of the amplitude $T_{ij}$ in the so-called \textit{second Riemann sheet}, obtained by using the following rotation of the loop function \cite{Montana:2017kjw}:
\begin{equation}
 G_{l}^\text{II}(\sqrt{s}+i\epsilon)=G_{l}(\sqrt{s}+i\epsilon)+i\frac{q_l}{4\pi\sqrt{2}}.
\end{equation}

Around the pole position $\sqrt{s_p}=M_p+i\Gamma_p/2$ the scattering amplitude can be approximated as
\begin{equation}
    T_{ij}(\sqrt{s})\simeq\frac{g_ig_j}{\sqrt{s}-\sqrt{s_p}},
\end{equation}
from which we can obtain the coupling constants, $g_i$, of the pole/resonance for all channels. 

In addition, we provide a measure of the content of the
$i^{\rm th}$ channel meson-baryon component in a given resonance. Let us note that the “compositeness” of resonances, defined as
 \begin{equation}
  X_i=  \left. g_i^2\displaystyle\frac{\partial G_i(\sqrt{s})}{\partial \sqrt{s}}\right|_{\sqrt{s}=\sqrt{s_p}}
  \end{equation}
in analogy to that of bound states, is a complex quantity that represents the integral of the $i^{\rm th}$ component of the wave-function squared, not the modulus squared (see Ref. \cite{Kinugawa:2024crb} for a summary of the present understanding of the compositeness in hadrons). Therefore, strictly speaking, the former definition cannot be interpreted as a probability, and some prescriptions have been given in the literature to connect this complex quantity to a real one, $\chi_i$, which provides a measure of the strength of the $i^{\rm th}$ channel within the resonance wave-function. We have taken $\chi_i=\left| X_i\right|$, which is the interpretation derived in Ref.~\cite{Guo:2015daa} and requires the resonance mass to be larger than the decaying channels. So, formally, our results for $\chi_i$ in Tables III, V, VI and VII should only have a meaningful probabilistic interpretation for channels below the quoted resonance mass.  An alternative definition is to associate $\chi_i$ to the real pat of $X_i$, as adopted in Ref.~\cite{Aceti:2014ala}, a prescription which ensures the fulfillment of the sum rule $\sum_{i} \chi_i + Z = 1$, where $Z$ is the amount of the genuine (non-molecular) component of the resonance. Other prescriptions for real valued compositeness in resonances can be found in \cite{Kamiya:2015aea,Sekihara:2015gvw}.

\section{Molecular resonances in S=-2 sector}
\subsection{Light sector}

We first discuss the results of the light channels of the pseudoscalar meson baryon interaction, where we employ the dimensional regularisation scheme to compute the loop function. The values of the unknown subtraction constants are determined by imposing the loop function to coincide, for a regularisation scale of $\mu=1000~{\rm MeV}$, with the cut-off loop function with $\Lambda=800~{\rm MeV}$ at the corresponding threshold.\footnote{Note that, in this light sector, the cut-off loop function employed to determine the subtraction constants is not that of Eq.~(\ref{eq:a(mu)}), but the alternative expression which is obtained when only the positive energy term of the baryon propagator is retained. This is the expression employed e.g.  in \cite{Oset:1997it}.}  We chose this value as it corresponds to the mass of the interchanged vector meson on the t-channel diagram which integrated out when we take the $t\to0$ limit. We will refer to this procedure as Model PB1. 

\begin{figure}[h!]
    \includegraphics[scale=0.35]{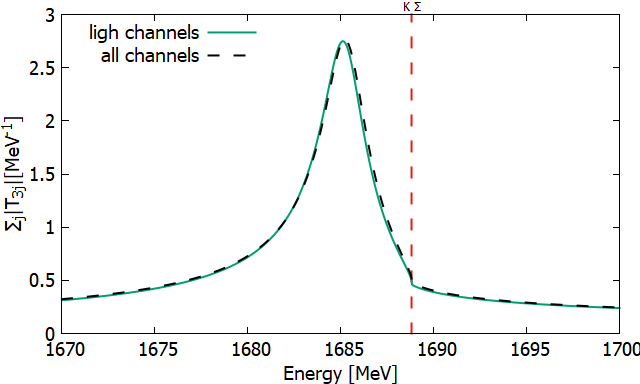}
    \caption{Sum over all $j$ channels of the modulus of the PB scattering amplitude, $|T_{3j}|$, obtained in Model PB1, as a function of the center-of-mass energy. The solid line shows the results when the heavy channels are neglected and the dashed one presents the results using all nine channels. The vertical coloured dashed lines represent the location of the thresholds of the different channels.}
    \label{fig:5vs9chan}
\end{figure}

Of course, first of all we have checked that our idea of separation between light and heavy channels works. As a typical example of the obtained effect, in Fig.  \ref{fig:5vs9chan} we show the sum  over all $j$ channels of the modulus of the PB scattering amplitude from the given channel $j$ into $\bar K \Sigma$, i.e. $\sum_j\, |T_{3j}|$.  As expected, the effect of heavy channels is negligible.

\begin{figure}[h!]
\subfloat[]{\includegraphics[width = 3in]{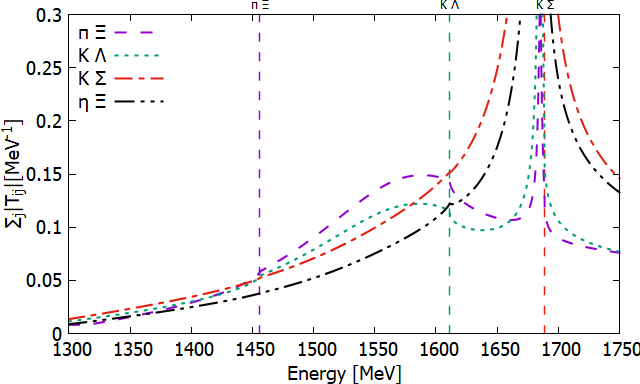}}\\ 
\subfloat[]{\includegraphics[width = 3in]{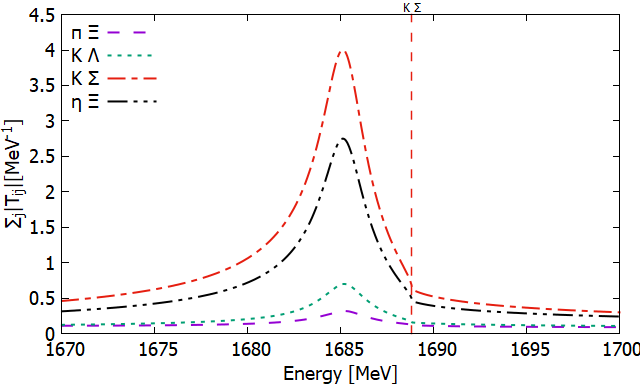}} 
\caption{(a): Sum over all $j$ channels of the modulus of the PB scattering amplitude, $|T_{ij}|$, for a fixed channel $i$, obtained in Model PB1 as a function of the center-of-mass energy. (b): Same as (a) but showing the region of the second pole with more details. The vertical dashed lines represent the location of the different channel thresholds.}
\label{fig:PBM1}
\end{figure}

The results of our calculations within Model PB1 are shown in Fig. \ref{fig:PBM1}, where we represent $\sum_j|T_{ij}|$ as a function of the c.m. energy for all different $i$ channels in different colours. These amplitudes show a broad peak (barely visible) at low energies and a narrower one right below the $\bar{K}\Sigma$ threshold. This behaviour is the reflection of two poles, the properties of which are shown in Table \ref{Tab:PBlight}. Note that these states have a well defined spin-parity of $J^\pi=\frac{1}{2}^-$, since they are formed with a scalar meson $(J^\pi=0^-)$ and a baryon $(J^\pi=\frac{1}{2}^+)$ interacting in s-wave.

 \begin{table}[hbt!]
  \begin{center}
    \begin{tabular}{ccc|cc|cc}
        \hline
        \multicolumn{7}{c}{$0^- \oplus \frac{1}{2}^+$ interaction in the $(I,S)=(\frac{1}{2},-2)$ sector}\\
        \hline
        \multicolumn{7}{c}{Model PB1}\\
        \hline
         $M(\rm MeV)$& & & \multicolumn{2}{c|}{$1570.53$} & \multicolumn{2}{c}{$1685.18$}\\
        $\Gamma({\rm MeV})$& & & \multicolumn{2}{c|}{$174.04$}   & \multicolumn{2}{c}{$2.28$}\\
        \hline
        & & & \multicolumn{2}{c|}{} & \multicolumn{2}{c}{}\\ [-3mm]
        & $a_i$ & $\Lambda~~~~$ & $~~~~|g_i|~~~$ & $ ~~~\chi_i~~~$ & $~~~|g_i|~~~$ & ~~~$\chi_i$~~~\\
         $\pi\Xi(1456)$         & $-1.51$ &  $800$ & $2.44$ & $0.428$ & $0.12$ & $0.001$\\
         $\bar{K}\Lambda(1611)$ & $-1.32$ &  $800$ & $1.91$ & $0.169$ & $0.27$ & $0.006$\\
         $\bar{K}\Sigma(1689)$  & $-1.27$ &  $800$ & $0.71$ & $0.016$ & $1.54$ & $0.815$\\
         $\eta\Xi(1866)$        & $-1.47$ &  $800$ & $0.49$ & $0.004$ & $1.06$ & $0.029$\\
        \hline
        \multicolumn{7}{c}{Model PB2}\\
        \hline
         $M(\rm MeV)$& & & \multicolumn{2}{c|}{$1600.55$} & \multicolumn{2}{c}{$1685.40$}\\
        $\Gamma({\rm MeV})$& & & \multicolumn{2}{c|}{$228.64$}   & \multicolumn{2}{c}{$2.60$}\\
        \hline
        & & & \multicolumn{2}{c|}{} & \multicolumn{2}{c}{}\\ [-3mm]
        & $a_i$ & $\Lambda$ & $|g_i|$ & $\chi_i$ & $|g_i|$ & $\chi_i$\\
         $\pi\Xi(1456)$         & $-1.30$ &  $630$ & $2.70$ & $0.501$ & $0.15$ & $0.001$\\
         $\bar{K}\Lambda(1611)$ & $-1.00$ &  $740$ & $2.21$ & $0.214$ & $0.27$ & $0.006$\\
         $\bar{K}\Sigma(1689)$  & $-1.27$ &  $800$ & $0.86$ & $0.025$ & $1.53$ & $0.822$\\
         $\eta\Xi(1866)$        & $-1.47$ &  $800$ & $0.42$ & $0.003$ & $1.02$ & $0.027$\\
    \end{tabular}
  \end{center}
    \caption{Position, subtraction constants, cut off, couplings and compositeness of the $\Xi$ states generated with the light channels of the PB interaction.}
    \label{Tab:PBlight}
\end{table}

We can observe that the lowest generated resonance couples strongly to the $\pi\Xi$ and $\bar{K}\Lambda$ channels, which are open for decay, what actually explains its rather large width. In the PDG \cite{PDG} one can find a state with similar mass, $\Xi(1620)$, which was seen decaying into the $\pi\Xi$ channel (see Table \ref{Tab:ExpTheo}). Even if our state is $50~{\rm MeV}$ lighter and its width is larger, we can see the potential of our model to generate a state in this region. 

As it can also be seen in Table \ref{Tab:PBlight} our heavier state mostly couples to the $\bar{K}\Sigma$ and  $\eta\Xi$ channels and it has a mass of $1685~{\rm MeV}$. According to the PDG there exists a $\Xi(1690)$ state at $1690\pm10~{\rm MeV}$ with a width of $20\pm15~{\rm MeV}$  (see Table \ref{Tab:ExpTheo}), which is compatible with our result.
{If this state was moved to a bit higher energy, the $\bar{K}\Sigma$ channel would be open for it to decay, and this would naturally lead to an increase of the width.

 \begin{table*}[hbt!]
  \begin{center}
    \begin{tabular}{ccccc|cc|cc}
        \hline
        \multicolumn{5}{c}{PDG data} & \multicolumn{2}{c}{Model PB1} & \multicolumn{2}{c}{Model PB2}\\
        \hline
         {\rm State} & {\rm Evidence} & $J^\pi$ & $M({\rm MeV})$ & $\Gamma({\rm MeV})$ &  $M({\rm MeV})$ & $\Gamma({\rm MeV})$ &  $M({\rm MeV})$ & $\Gamma({\rm MeV})$\\
        \hline
        $\Xi(1620)$ & ** & - & $\sim 1620$ & $32^{+8}_{-9}$ & $1570.53$ & $174.53$ & $1600.55$ & $228.65$ \\
        $\Xi(1690)$ & *** & - & $1690\pm10$ & $20\pm 15$ & $1685.18$ & $2.28$ & $1685.40$ & $2.60$ \\
        \hline
        \multicolumn{5}{c}{PDG} & \multicolumn{2}{c}{Model VB2} & \multicolumn{2}{c}{Model VB3}\\
        \hline
        $\Xi(1820)$ & *** & $\frac{3}{2}^-$ & $1823\pm5$ & $24^{+15}_{-10}$ & $1823.12$ & $0.66$ & $-$ & $-$ \\
        \multirow{2}{*}{$\Xi(1950)$} & \multirow{2}{*}{***} & \multirow{2}{*}{-} & \multirow{2}{*}{$1950\pm15$} & \multirow{2}{*}{$60\pm20$} & \multirow{2}{*}{$1949.23$} & \multirow{2}{*}{$0.232$} & $1935.99$ & $8.29$ \\
        & & & & & & & $1964.09$ & $4.92$
    \end{tabular}
  \end{center}
    \caption{Experimental information for the $\Xi(1620)$, $\Xi(1690)$, $\Xi(1820)$, $\Xi(1950)$ resonances and the mass and width for the states generated with four different models.}
    \label{Tab:ExpTheo}
\end{table*}

Looking at the results, it is natural to think that we can modify the values of the subtraction constants within a reasonable range  in order to accommodate the generated states to the experimental data. Therefore, we relax the condition of the loop function matching the cut-off one with $\Lambda=800~{\rm MeV}$ at the corresponding threshold \cite{Montana:2017kjw}. By doing that we define a new model, called PB2. The results of new calculations within Model PB2 are shown in Fig. \ref{fig:PBM2} and in Table \ref{Tab:PBlight}. We can see that the lighter resonance now appears at higher mass and it is much wider, while the heavier resonance is generated with a similar mass and with a slightly larger width. Looking at Table \ref{Tab:ExpTheo} we can see that now the mass of the light state is closer to the experimental energy of the $\Xi(1620)$. In this model, none of the equivalent cut-off values for the subtractions constants employed are smaller than $\Lambda=630~{\rm MeV}$, which is a reasonable lower limit value.

\begin{figure}[h!]
\subfloat[]{\includegraphics[width = 3in]{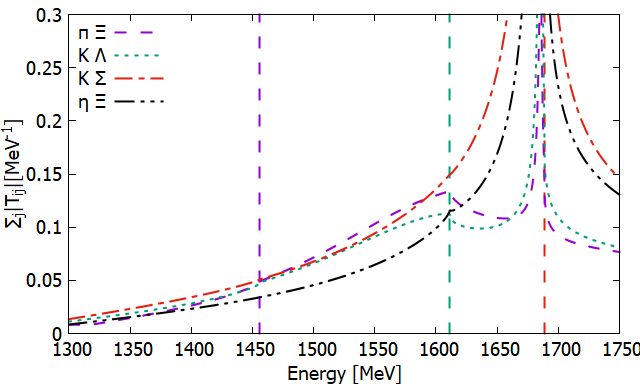}}
\caption{Sum over all $j$ channels of the modulus of the PB scattering amplitude, $|T_{ij}|$, for a fixed channel $i$, obtained in Model PB2 as a function of the center-of-mass energy. The vertical dashed lines represent the location of the different channel thresholds.}
\label{fig:PBM2}
\end{figure}

Reviewing the literature \cite{Sekihara:2015qqa, Kaiser:1995eg, Oset:1997it, Ramos:2002xh, Garcia-Recio:2003ejq, Gamermann:2011mq, Nishibuchi:2023acl, Nishibuchi:2022zfo, Nishibuchi:2023nvi}
one can note that all the models based on just a leading interaction term in the potential, the so called Weinberg-Tomozawa term, see Eq.~(\ref{eq:Vij}), can dynamically generate two poles, the features of which cannot reproduce
the experimental masses and widths of the $\Xi(1620)$ and the $\Xi(1690)$ at
the same time.  Furthermore, the
width of the $\Xi(1690)$ remains always rather small, around a few MeV. Consistently, our calculations also show all these features.
In some cases   
the $\Xi(1620)$ resonance can be reproduced very well \cite{Nishibuchi:2023acl, Nishibuchi:2022zfo, Nishibuchi:2023nvi}, but only allowing some of the subtracting constants to take unnatural-sized values.

These limitations can be overcome if one also takes into consideration the s- and u-channel Born diagrams and
the NLO contribution, as it was done in \cite{Feijoo:2023wua,Magas:2024mba,Sarti:2023wlg}, where the properties of both $\Xi(1620)$ and $\Xi(1690)$ resonances are well reproduced by the model. 

We proceed now to discuss the results for the vector meson baryon interaction, where we construct the model VB1 in a similar way as to the model PB1, i. e. using a regularisation scale of $\mu=1000~{\rm MeV}$ and requesting that our loop function, regularized via dimensional regularization, for each channel should have the same value at the threshold as that computed with cut-off $\Lambda=800~{\rm MeV}$. Using this model we produce the amplitude displayed in Fig. \ref{fig:VBM1}, which has two poles with properties listed in Table \ref{Tab:VBlight}. Note that each of these states forms a degenerate spin-parity doublet with $J^\pi=\frac{1}{2}^-$, $J^\pi=\frac{3}{2}^-$, since they are produced from the interaction of a vector meson $(J^\pi=1^-)$ an a baryon $(J^\pi=\frac{1}{2}^+)$ in s-wave.

\begin{figure}[h!]
    \includegraphics[scale=0.37]{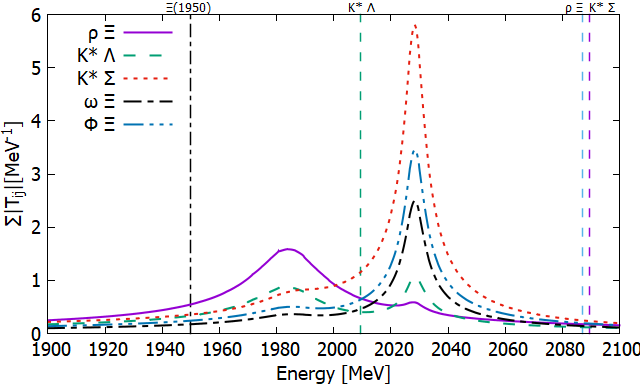}
    \caption{Sum over all $j$ channels of the modulus of the VB scattering amplitude, $|T_{ij}|$ for a fixed channel $i$, obtained in Model VB1 as a function of the center-of-mass energy. The vertical coloured dashed lines represent the location of the different channel thresholds and the black dot-dashed one denotes the mass position of the $\Xi(1950)$.}
    \label{fig:VBM1}
\end{figure}

 \begin{table}[hbt!]
  \begin{center}
    \begin{tabular}{ccc|cc|cc}
        \hline
        \multicolumn{7}{c}{$1^- \oplus \frac{1}{2}^+$ interaction in the $(I,S)=(\frac{1}{2},-2)$ sector}\\
        \hline
        \multicolumn{7}{c}{Model VB1}\\
        \hline
         M(Mev)& & & \multicolumn{2}{c|}{1985.95} & \multicolumn{2}{c}{2028.40}\\
        $\Gamma({\rm MeV})$& & & \multicolumn{2}{c|}{29.23}   & \multicolumn{2}{c}{6.86}\\
        \hline
        & & & \multicolumn{2}{c|}{} & \multicolumn{2}{c}{}\\ [-3mm]
        & $a_i$ & $\Lambda$ & $|g_i|$ & $ \chi_i$ & $|g_i|$ & $\chi_i$\\
         $\rho\Xi(2089)$     & $-1.50$ &  800 & 3.39 & 0.610 & 0.24 & 0.004\\
         $\bar{K}^*\Lambda(2010)$   & $-1.42$ &  800 & 1.82 & 0.358 & 0.52 & 0.0490\\
         $\bar{K}^*\Sigma(2087)$  & $-1.46$ &  800 & 0.98 & 0.047 & 2.97 & 0.629\\
         $\omega\Xi(2101)$ & $-1.50$ &  800 & 0.54 & 0.014 & 1.30 & 0.114\\
         $\phi\Xi(2338)$   & $-1.57$ &  800 & 0.75 & 0.012 & 1.80 & 0.077\\
        \hline
        \multicolumn{7}{c}{Model VB2}\\
        \hline
         M(Mev)& & & \multicolumn{2}{c|}{1823.12} & \multicolumn{2}{c}{1949.23}\\
        $\Gamma({\rm MeV})$& & & \multicolumn{2}{c|}{0.66}   & \multicolumn{2}{c}{0.232}\\
        \hline
        & & & \multicolumn{2}{c|}{} & \multicolumn{2}{c}{}\\ [-3mm]
        & $~~a_i~~$ & $~~\Lambda~~$ & $~~~|g_i|~~~$ & $~~~ \chi_i~~~$ & $~~~|g_i|~~~$ & $~~~\chi_i~~~$\\
         $\rho\Xi(2089)$          & $-2.19$ & 1440 & 3.48 & 0.323 & 0.16 & 0.001\\
         $\bar{K}^*\Lambda(2010)$ & $-2.00$ & 1300 & 2.07 & 0.131 & 0.66 & 0.035\\
         $\bar{K}^*\Sigma(2087)$  & $-1.87$ & 1140 & 1.16 & 0.033 & 3.15 & 0.509\\
         $\omega\Xi(2101)$        & $-1.50$ &  800 & 0.41 & 0.004 & 1.43 & 0.100\\
         $\phi\Xi(2338)$          & $-1.80$ &  990 & 0.56 & 0.005 & 1.98 & 0.100\\
        \hline
        \multicolumn{7}{c}{Model VB3}\\
        \hline
         M(Mev)& & & \multicolumn{2}{c|}{1935.99} & \multicolumn{2}{c}{1964.09}\\
        $\Gamma({\rm MeV})$& & & \multicolumn{2}{c|}{8.29}   & \multicolumn{2}{c}{4.92}\\
        \hline
        & & & \multicolumn{2}{c|}{} & \multicolumn{2}{c}{}\\ [-3mm]
        & $a_i$ & $\Lambda$ & $|g_i|$ & $ \chi_i$ & $|g_i|$ & $\chi_i$\\
         $\rho\Xi(2089)$          & $-1.74$ &  990 & 2.88 & 0.335 & 1.43 & 0.096\\
         $\bar{K}^*\Lambda(2010)$ & $-1.42$ &  800 & 1.41 & 0.116 & 1.53 & 0.186\\
         $\bar{K}^*\Sigma(2087)$  & $-1.82$ & 1090 & 2.58 & 0.250 & 2.66 & 0.309\\
         $\omega\Xi(2101)$        & $-1.50$ &  800 & 0.41 & 0.006 & 1.45 & 0.092\\
         $\phi\Xi(2338)$          & $-1.85$ & 1040 & 0.56 & 0.006 & 2.00 & 0.081\\
    \end{tabular}
  \end{center}
    \caption{Position, subtraction constants, cut off, couplings and compositeness of the $\Xi$ states generated with the light channels of the VB interaction.}
    \label{Tab:VBlight}
\end{table}

As it can be seen in Table \ref{Tab:VBlight}, the lighter state couples strongly to the $\rho\Xi$ channel. Note that, although the mass of this state is lower than all the channel thresholds, it has a non-zero width. This is because we take into account the width of the $\rho$ and the  $\bar{K}^*$ in the loop function. We can see that the higher energy state  couples strongly to the $\bar{K}^*\Sigma$ channel. In the PDG compilation one finds two states in this region, $\Xi(1820)$ and  $\Xi(1950)$ (see Table \ref{Tab:ExpTheo}). The first state has a known spin-parity $J^\pi=\frac{3}{2}^-$, which is compatible with our model.

\begin{figure}[h!]
    \includegraphics[scale=0.37]{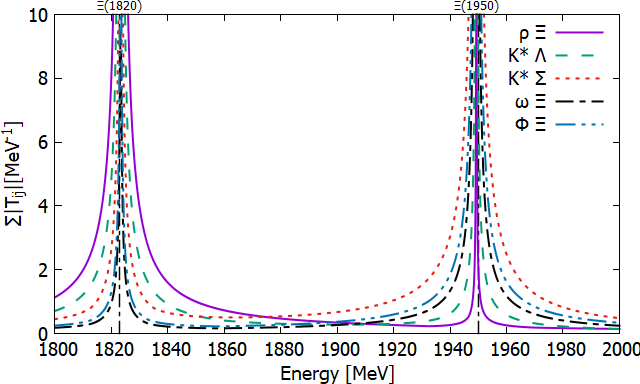}
    \caption{Sum over all $j$ channels of the modulus of the VB scattering amplitude, $|T_{ij}|$ for a fixed channel $i$, obtained in Model VB2 as a function of the center-of-mass energy. The vertical black dot-dashed lines represent the mass position of the $\Xi(1820)$ and the $\Xi(1950)$.}
    \label{fig:VBM2}
\end{figure}

Similarly to what we did in PB case, we will modify our subtraction constants within a reasonable range trying to adjust our results to the experimental data. In this way, we build model VB2, which produces two resonances at energies compatible with the experimental masses of $\Xi(1829)$ and $\Xi(1950)$ (see Fig. \ref{fig:VBM2} and Table \ref{Tab:VBlight}). We note that the values of the equivalent cut-offs for the subtractions constants employed are not larger than $\Lambda=1500~{\rm MeV}$, which is a reasonable upper limit value.  However, the widths of these obtained resonances are much smaller than the experimental ones. This problem might potentially be solved if we implemented the coupling of the VB channels with the PB ones, as this would open more light channels in which these resonances could decay, thereby increasing their width. Such a coupling was studied in \cite{Garzon:2012np}, via pseudoscalar exchange plus a contact (or Kroll-Ruderman) term (the latter turning out to be more important), and indeed it generated a moderate increase of the width for the lower lying resonance.

\begin{figure}[h!]
    \includegraphics[scale=0.37]{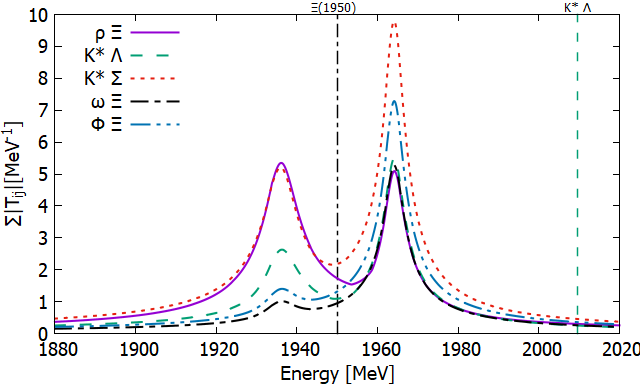}
    \caption{Sum over all j channels of the modulus of the VB scattering amplitude, $|T_{ij}|$ for a fixed channel $i$, obtained in Model VB3. The vertical coloured dashed lines represent the location of the thresholds of the different channels and the black one represents the mass position of the $\Xi(1950)$.}
    \label{fig:VBM3}
\end{figure}

Another possibility to adjust our model to the experimental data is to generate both resonances in the energy region where the $\Xi(1950)$ appears. As described in the PDG \cite{PDG} this state could be representing more than one $\Xi$ resonance, eventually merging into an apparent wide peak. With such a motivation we generated model VB3, which, as can be seen in Fig. \ref{fig:VBM3} and Table \ref{Tab:VBlight}, produces two states close to the experimental position of the $\Xi(1950)$. Note that this model required the modification of only three subtraction constants and the equivalent cut-off values are not larger than $1100~{\rm MeV}$ for all the channels. If our resonances were a bit wider, which could be achieved by coupling our VB channels with the PB ones, as discussed above, they would produce just one, but rather wide, peak in the spectrum.  

\subsection{Heavy sector} \label{sec:ResHeavy}

\begin{figure}[h!]
    \includegraphics[scale=0.37]{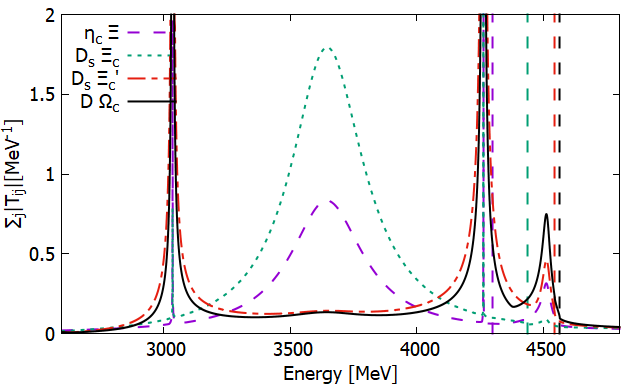}
    \caption{Sum over all $j$ channels of the modulus of the PB scattering amplitude, $|T_{ij}|$ for a fixed channel $i$, obtained with the dimensional regularization scheme for the loop function, as a function of the center-of-mass energy. The vertical dashed lines represent the location of the different channel thresholds.}
    \label{fig:HeavydrPB}
\end{figure}

Let us now discuss the results in the heavy sector, where there are only four channels. Some results have already been published in Ref.~\cite{Marse-Valera:2022khy} in a rather brief form. In this section we will provide more details of the calculations and and some additional results. 

Similarly to the previous section, we first set the regularization scale to $\mu=1000~{\rm MeV}$ and the subtraction constants are evaluated by mapping, at the threshold of each channel, the dimensional regularized loop function to the value obtained with the cut-off expression employing $\Lambda=800~{\rm MeV}$. In such a way four peaks are generated, as we can see in Fig. \ref{fig:HeavydrPB}.

\begin{figure*}[t]
    \includegraphics[scale=0.45]{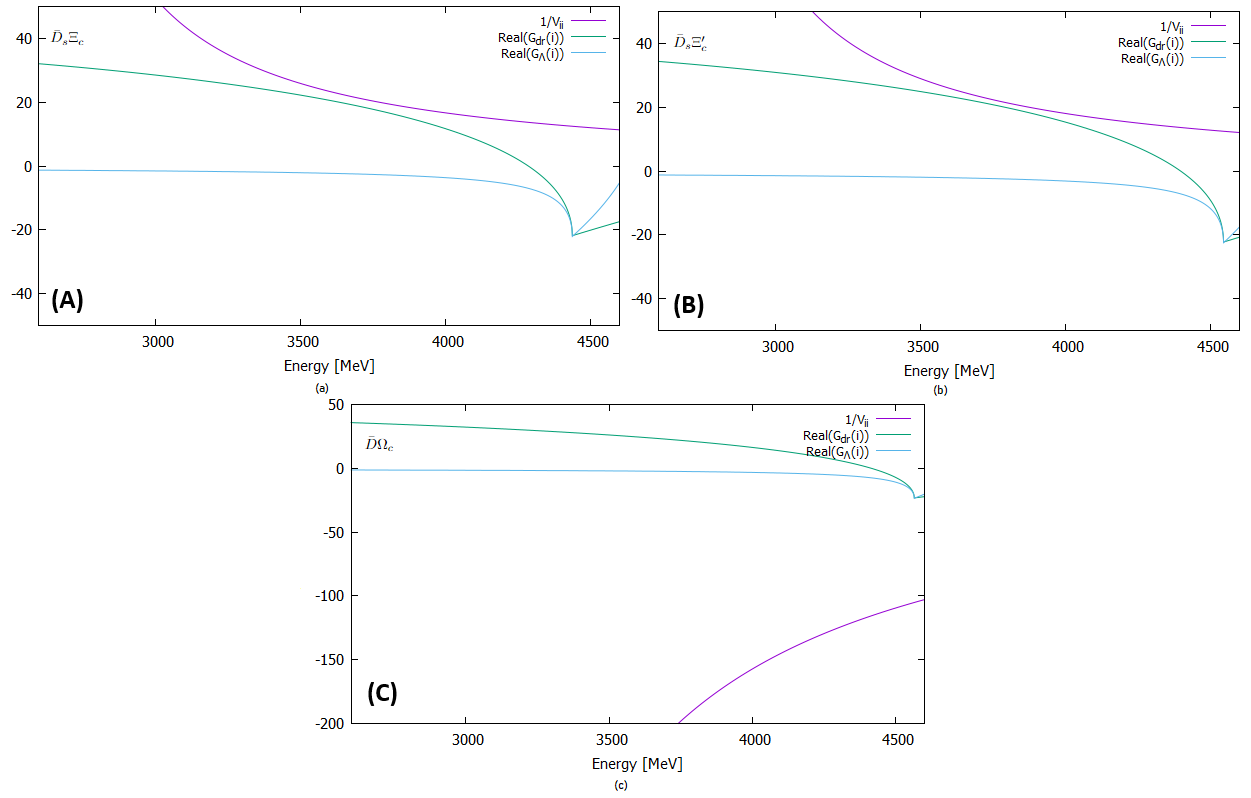}
    \caption{The purple line represents the inverse of the diagonal term of the potential, the green line shows the loop function of the same channel, computed using a dimensional regularization scheme, and the blue line is similar to the green one, but computed using a cut-off scheme. In (a), (b) and (c) we show the $\bar{D}_s\Xi_c$,  $\bar{D}_s\Xi_c'$ and $\bar{D}\Omega_c$ channel, respectively. 
    }
    \label{fig:inVREalG}
\end{figure*}

Analyzing these results wee observe that the first three peaks couple very strongly to $\bar{D}_s~\Xi'_c$ and/or to $\bar{D}_s~\Xi_c$, which is surprising since the diagonal elements of the potential for these channels are
repulsive ($C_{ii} < 0$ in Table \ref{Tab:CijPB}). This might be an indication that these peaks
might be mathematical artifacts of the dimensional regularization scheme. To illustrate the problem, we present in Figs. \ref{fig:inVREalG} the inverse of the diagonal potential and the real part of the loop function, the latter calculated using both the dimensional regularization and the cut-off schemes, for the three heavier channels.
Note that, according to Eq.~(\ref{eq:Tij2}), a pole is generated when $VG\sim1$.  As we can see in Figs. \ref{fig:inVREalG} (a) and (b) the inverse of the potential and the real part of the dimensional regularized loop function almost intersect in a region where the potential is repulsive. Thus, these states, which mostly couple to $\bar{D}\Xi_c$ and $\bar{D}\Xi_c'$, are only mathematical solutions of our equation, but do not represent any physical state. Fig. \ref{fig:inVREalG}  also shows such fake poles could be avoided by employing the cut-off scheme to compute the loop functions.

\begin{figure}[h!]
    \includegraphics[scale=0.37]{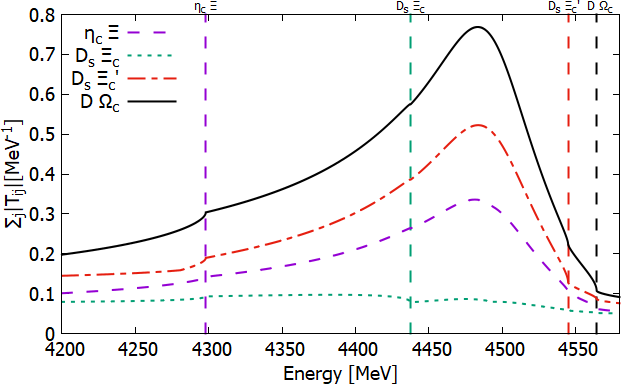}
    \caption{Sum over all $j$ channels of the modulus of the PB scattering amplitude, $|T_{ij}|$, obtained with the cut-off model to compute the loop function, as a function of the center-of-mass energy. The vertical dashed lines represent the location of the different channel thresholds.}
    \label{fig:cutoffPB}
\end{figure}

As can be seen in Fig. \ref{fig:cutoffPB}, the fake poles disappear when we use a cut-off scheme with $\Lambda=800~{\rm MeV}$ and, as expected, only the pole at higher energy remains. This one strongly couples to the $\bar{D}\Omega_c$ and $\bar{D}_s\Xi_c'$ channels (see Table \ref{tab:PBHigh}). This state, which would be a pentaquark with double strangeness ($P_{\Psi_{ss}}^\Xi$) has a well defined spin-parity $J^\pi=\frac{1}{2}^-$ as it is generated from the PB interaction in s-wave.

 \begin{table}[hbt!]
  \begin{center}
    \begin{tabular}{ccccc}
        \hline
        \multicolumn{5}{c}{$0^- \oplus \frac{1}{2}^+$ interaction in the $(I,S)=(0,-2)$ sector}\\
        \hline
         M(Mev)& &  \multicolumn{3}{c}{4493.35}\\ [-1mm]
        $\Gamma({\rm MeV})$ & & \multicolumn{3}{c}{73.67}\\
        \hline
        & $\Lambda$ & $g$ & $|g_i|$ & $ \chi_i$ \\
         $\eta_c\Xi(4302)$            &  800 & $-1.60+0.34i$ &1.63 & 0.220\\
         $\bar{D}_s\Xi_c(4437)$     &  800 & $-0.17+0.27i$ & 0.32 & 0.019\\
         $\bar{D}_s\Xi_c'(4545)$    &  800 & $ -2.41+0.58i$ & 2.48 & 0.398\\
         $\bar{D}\Omega_c(4564)$    &  800 & $ 3.59-0.77i$ & 3.67 & 0.711\\
    \end{tabular}
  \end{center}
    \caption{Position, cut off, couplings and compositeness of the $\Xi$ state generated with the heavy channels of the PB interaction.}
    \label{tab:PBHigh}
\end{table}

It is very interesting to look for the origin of this resonance. Checking Table \ref{Tab:CijPB} we can see that only diagonal $\bar{D}\Omega_c$ term is positive (i.e. attractive in our notation) although with a rather small
strength $(C_{99}=1/9)$. Such an attraction is not sufficient by itself to give rise to a bound state, a fact that we checked upon performing an uncoupled calculation. Thus, as already noted in our previous work \cite{Marse-Valera:2022khy}, this resonance is generated in a rather unusual way from the non-diagonal terms of the interaction. Indeed, the $\bar{D}\Omega_c - \bar{D}_s\Xi_c'$ interaction is strong enough ($C_{98}=-\sqrt{2}$), to produce a bound state via the coupled channels. To check this point, we added a $\gamma$ factor to this coupling, namely $C_{89}\to C_{89}'=-\gamma\sqrt{2}$,  which we varied to study the effect of this off-diagonal term in detail, as can be seen in Fig. \ref{fig:cutoffPBgam}. We observe that the resonance is not generated for values of $\gamma \lesssim 0.5$. 

\begin{figure}[h!]
    \includegraphics[scale=0.37]{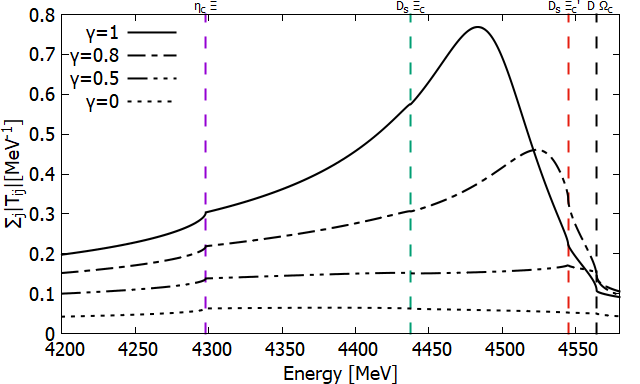}
    \caption{Sum over all $j$ channels of the modulus of the PB scattering amplitude, $|T_{ij}|$, for $i=\bar{D}\Omega_c$ and for values of $\gamma$ ($C_{89}\to C_{89}'=\gamma C_{89}$), as a function of the center-of-mass energy. The loop function is calculated with a cut-off scheme as in Fig. \ref{fig:cutoffPB}. The vertical dashed lines represent the location of the different channel thresholds.}
    \label{fig:cutoffPBgam}
\end{figure}

\begin{figure}[h!]
    \includegraphics[scale=0.37]{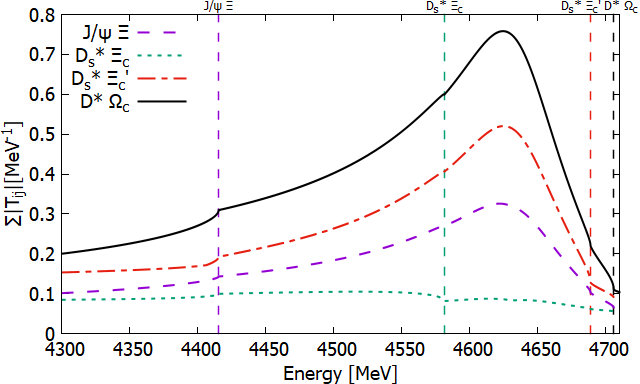}
    \caption{Sum over all $j$ channels of the modulus of the VB scattering amplitude, $|T_{ij}|$ for a fixed channel $i$, obtained using the cut-off scheme to compute the loop function, as a function of the center-of-mass energy.  The vertical dashed lines represent the location of the different channel thresholds.}
    \label{fig:cutoffVB}
\end{figure} 

We now present our results in the heavy VB sector. In the similar way as in the PB sector, the use of dimension regularization for the loop function leads to some fake nonphysical poles, which have been eliminated by the use of the cut-off method. Finally, only one resonance state remains, which couples strongly to $\bar{D}^*\Omega_c$ and  $\bar{D}_s^*\Xi_c'$, see Fig. \ref{fig:cutoffVB} and Table \ref{tab:VBHigh}. Similarly to the PB sector this state can be formed due to the large value of the $C_{89}$ coefficient.
Note that this state is degenerate in spin-parity, can be either $J^P=\frac{1}{2}^-$ or $J^P=\frac{3}{2}^-$, since we have a vector-meson $(J^P=1^-)$ interacting with a baryon $(J^P=\frac{1}{2}^+)$ in s-wave.

 \begin{table}[hbt!]
  \begin{center}
    \begin{tabular}{ccccc}
        \hline
        \multicolumn{5}{c}{$1^- \oplus \frac{1}{2}^+$ interaction in the $(I,S)=(0,-2)$ sector}\\
        \hline
         M(Mev)& &  \multicolumn{3}{c}{4633.38}\\
        $\Gamma({\rm MeV})$ & & \multicolumn{3}{c}{79.58}\\
        \hline
        & $\Lambda$ & $g$ & $|g_i|$ & $ \chi_i$ \\
         $J/\psi\Xi(4415)$            &  800 & $-1.62+0.38i$  & 1.66 & 0.252\\
         $\bar{D}_s^*\Xi_c(4581)$     &  800 & $-0.143+0.32i$ & 0.34 & 0.022\\
         $\bar{D}_s^*\Xi_c'(4689)$    &  800 & $-2.49+0.67i$  & 2.58 & 0.406\\
         $\bar{D}^*\Omega_c(4706)$    &  800 & $3.67+0.89i$   & 3.78 & 0.740\\
    \end{tabular}
  \end{center}
    \caption{Position, cut off, couplings and compositeness of the $\Xi$ states generated with the heavy channels of the VB interaction. }
        \label{tab:VBHigh}
\end{table}


We also have checked that the variations of cut off value, of $\kappa_c$  and $\kappa_{cc}$ parameters (related to explicit violation of SU(4) symmetry) lead to some variations of the resonance position and properties, but the appearance of a pole is a robust outcome in all calculations both for PB and VB cases \cite{Marse-Valera:2022khy}. 

Thus we have shown that a simple model with realistic parameters does generate three pentaquarks with double strangeness and hidden charm.  Shortly after our first results were published \cite{Marse-Valera:2022khy}, another study performed with a similar model \cite{Roca:2024nsi} predicted pentaquarks in the same mass region, but with much smaller widths.  As explained in Ref.~\cite{Roca:2024nsi}, the main reason for this difference is the relative factor of $-1/\sqrt{3}$ factor multiplying their
$\bar{D}_s\Xi_c'\to \eta_c\Xi$ and $\bar{D}\Omega_c \to \eta_c\Xi$ transitions. Indeed, we have checked that implementing theses factors our model generates resonances practically at the same place as before but with a smaller width of about $22~{\rm MeV}$. This width is still about twice as that found in \cite{Roca:2024nsi} because our model also contemplates the additional  $\bar{D}_s\Xi_c$ channel with a sizable coupling to $\eta_c\Xi$, the primary final decaying state. 

\section{The  $\Xi_b \to  J/\Psi  \phi \Xi$ decay: formalism}

\subsection{Dalitz Plot}
\label{dalitz}
As mentioned earlier, the authors of Ref. \cite{Oset:2024fbk} suggest the measurement of the $\Xi_b^0\to \eta \eta_c \Xi^0$ and $\Omega_b^-\to K^- \eta_c \Xi^0$ decays in order to observe the $P_{css}$ state with $I(J^P)=\frac{1}{2}(\frac{1}{2}^-)$, generated from a pseudoscalar-baryon interaction, in the invariant mass of the $\eta_c\Xi^0$ pairs. 

In the present work we investigate the $\Xi_b\to\Xi~J/\psi~\phi$ decay as a possible process to look for a double strangeness pentaquark, coupling strongly to a vector meson-baryon pair, in the $J/\psi \Xi$ invariant mass. We will first inspect the Dalitz plots representing the regions of two-body invariant mass distributions of the three possible pairs of particles $(J/\psi\phi,~J/\psi\Xi,~\phi\Xi)$ to show that there exists a good possibility to detect not only pentaquarks $P_{css}$, but other exotic hadrons as well. 


  \begin{figure*}[t]
     \includegraphics[scale=0.40]{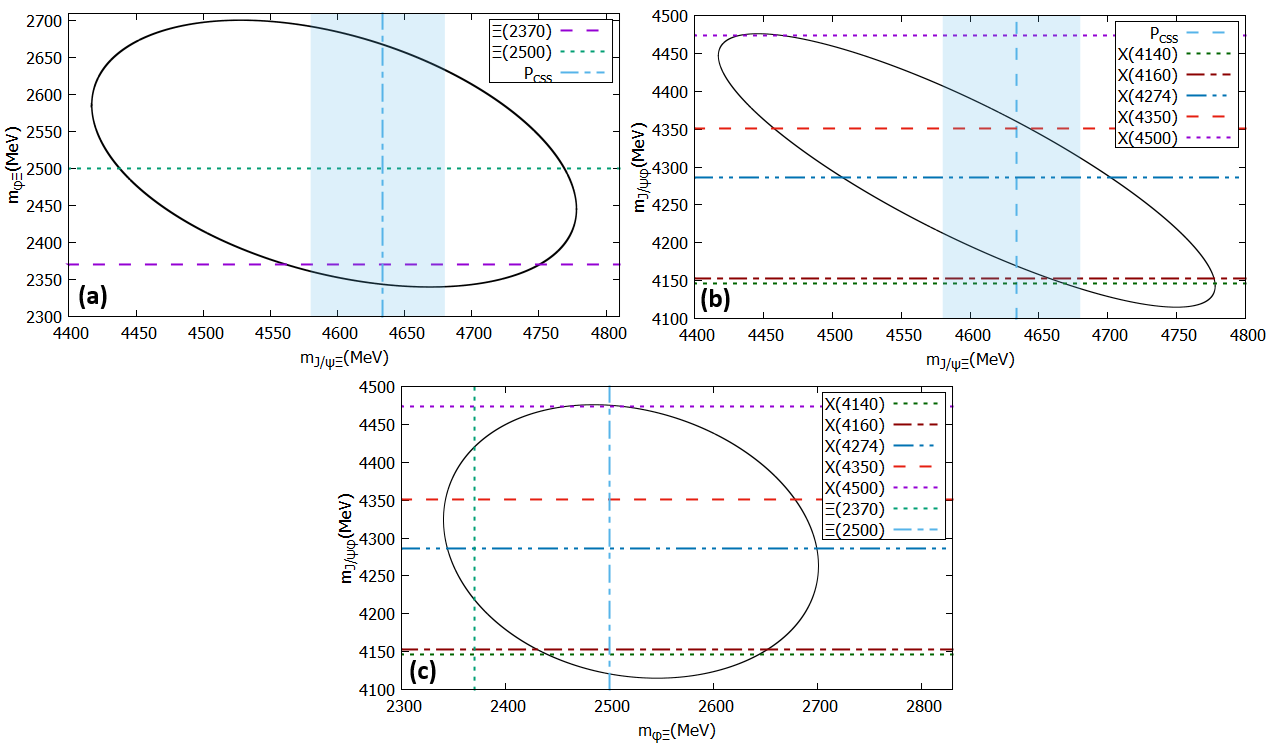}
     \caption{The solid lines represent the Dalitz plot for $M_{\phi\Xi}$ vs $M_{J/\psi\Xi}$ (a), $M_{J/\psi\phi}$ vs $M_{J/\psi\Xi}$ (b) and $M_{J/\psi\phi}$ vs $M_{\phi\Xi}$ (c). The dashed lines show the mass position for different resonances that couple to the corresponding channel. The shadowed zone is the variation we considered in this work for the pentaquark position.}
     \label{Fig:Dalitz}
 \end{figure*}

In the first two Dalitz plots represented in Fig. \ref{Fig:Dalitz} (panels (a) and (b)) the dashed vertical line corresponds to the mass of the $S=-2$ pentaquark predicted by our $VB$ model ($4633$ MeV) and the shadowed zone shows the variations of its mass of $\simeq\pm50~{\rm MeV}$ accounting for theoretical uncertainties. It is clear that the invariant mass range of the $J/\psi \Xi$ pairs covers amply the region where our double strangeness pentaquark is predicted and it should therefore leave a signal in their spectrum.  

Other examples of exotic states are the $X$ meson resonances, which can be observed in the ${J/\psi\phi}$ invariant mass spectrum. 
The Dalitz plots (b) and (c) of Fig. \ref{Fig:Dalitz} show that five $X$ resonances,  the $X(4140)$, $X(4160)$, $X(4274)$, $X(4550)$ and the $X(4700)$, lie inside the permitted kinematical region. In spite of the fact that all five states may be detectable, we will only focus on the first two states, as they have been the object of some controversy. Indeed, 
there is still a discussion of the nature and properties of the $X(4140)$ and $X(4160)$ states. In 2008, the Belle collaboration reported the existence of the $X(4160)$ \cite{Belle:2007woe} in the $e^-e^+\to J/\psi D^*\Bar{D}^*$ reaction and, during the following years, some groups reported the existence of a narrow $X(4140)$ with a width around $19~{\rm MeV}$ \cite{CDF:2009jgo,CDF:2011pep,LHCb:2012wyi,CMS:2013jru,D0:2013jvp,D0:2015nxw}. Despite of these results, a more recent measurement of the $B^+\to J/\psi\phi K^+$ decay from the LHCb collaboration \cite{LHCb:2016axx} obtained a width of $83~{\rm MeV}$ for the $X(4140)$, which is pretty large compared with the previously studies. In that work other states that couple to $J/\psi\phi$ were also reported such as the $X(4274)$, $X(4550)$ and the $X(4700)$, but the $X(4160)$, which is the state that was previously seen in 2008, was not observed. It might be that the reason of this large width for $X(4140)$ is related to the fact that the two neighboring states (the narrow $X(4140)$ state plus a wider $X(4160)$ resonance) were fitted together. In this work we will study the interplay between the $X(4140)$  and the $X(4160)$ resonances and will try to find an observable that permits us to discriminate whether the truly nature of this state is only one wide $X(4140)$ resonance or it is the combination of a narrow $X(4140)$ state plus a $X(4160)$ one.

Among the theoretical works studying these $X$ states, some groups identified the $X(4140)$ as a molecular $D_s^*\Bar{D}_s^*$ bound state with quantum numbers $0^{++}$ and $2^{++}$ \cite{Liu:2009ei,Branz:2009yt,Chen:2015fdn}. These studies did not take into account the coupling to light meson states, thus producing a small width for the state, hence associating it to the $X(4140)$. The contribution of the light meson channels were included in Ref. \cite{Molina:2009ct}, which generated a $2^{++}$ $X$ state at $4169~{\rm MeV}$ with a width of $139~{\rm MeV}$, coupling strongly to to $D_s^*\Bar{D}_s^*$, and being associated to the $X(4160)$. In the end, the quantum numbers of the $X(4140)$ were measured, obtaining $J^{PC}=1^{++}$ \cite{PDG}, therefore the models that predict this state as a $D_s^*\Bar{D}_s^*$ molecule are no longer supported.

Focusing on the $\Xi_b\to\Xi X^*\to\Xi~J/\psi~\phi$ decay in the present work, apart from employing it to search for the double strangeness pentaquark, we also have a chance to study the interplay between the $X(4140)$ and $X(4160)$ resonances. Similarly to what was done in Refs. \cite{Magas:2020zuo,Wang:2017mrt}, we will use two independent primary decay mechanisms, one for the $X(4160)$ and another for the $X(4140)$. Since the $X(4140)$ resonance is not generated by any meson-meson interaction model, we assume for it a Breit-Wigner shape whose parameters are fitted to experimental data. In contrast, for the $X(4160)$ we will employ the results of a molecular model \cite{Molina:2009ct},  which reproduces rather well its mass and width and also predicts a strong coupling to $D_s^*\bar{D}_s^*$ channel.

\subsection{Primary decay} \label{sec:PrimDecay}

\begin{figure}[h!]
     \includegraphics[scale=0.40]{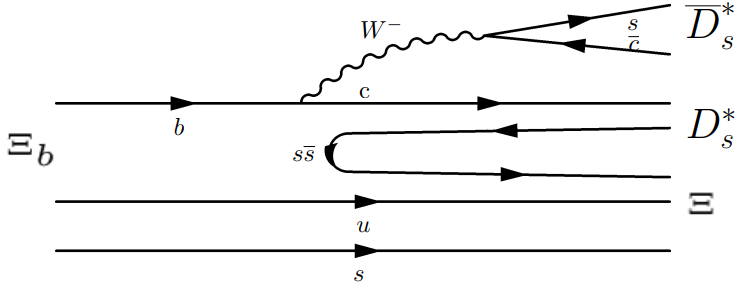}
     \caption{Microscopic quark level $\Xi_b\to\Xi~D_s^*~\Bar{D}_s^*$ transition, through external emission.}
     \label{Fig:DD}
 \end{figure}

  \begin{figure}[h!]
     \includegraphics[scale=0.40]{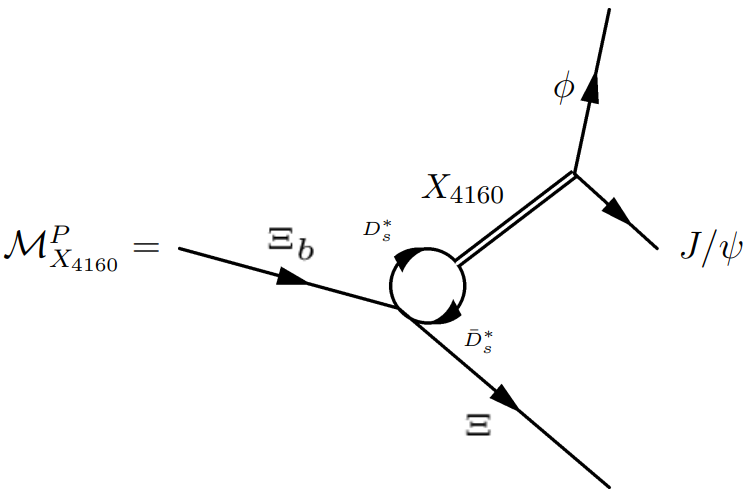}
     \caption{Mechanism for the $\Xi_b\to J/\psi~\phi~\Xi$ decay involving the $X(4160)$ resonance.}
     \label{Fig:X4160}
 \end{figure}
 
The dominant process for the $\Xi_b$ decay at the quark level proceeds via the external emission mechanism depicted in Fig. \ref{Fig:DD}. Since the final state involves $J/\psi \,\phi$ pairs,  the $D_s^* \, \Bar{D}_s^*$ lines have to be connected to form a loop, as shown in Fig. \ref{Fig:X4160},
allowing multiple coupled channel interactions and generating dynamically the $X(4160)$, which then decays into $J/\psi$ and $\phi$.
According to Ref. \cite{Molina:2009ct} the $X(4160)$ couples with maximum strength to the $D_s^* \, \Bar{D}_s^*$ channel, but its coupling to the $J/\psi \, \phi$ channel is also significant, what also increases the dominance of the process depicted in Fig. \ref{Fig:X4160}. On the other hand, if we draw a diagram which would be responsible even for a direct $J/\psi \, \phi \Xi$ decay,  at the microscopic quark level this reaction  proceeds via the internal conversion process, shown in Fig. \ref{int_con}, which is strongly penalized by color factors \cite{color_fac} and therefore it will be neglected in this work, similarly to what was done in Ref. \cite{Magas:2020zuo,Wang:2017mrt}.

\begin{figure}
    \includegraphics[scale=0.34]{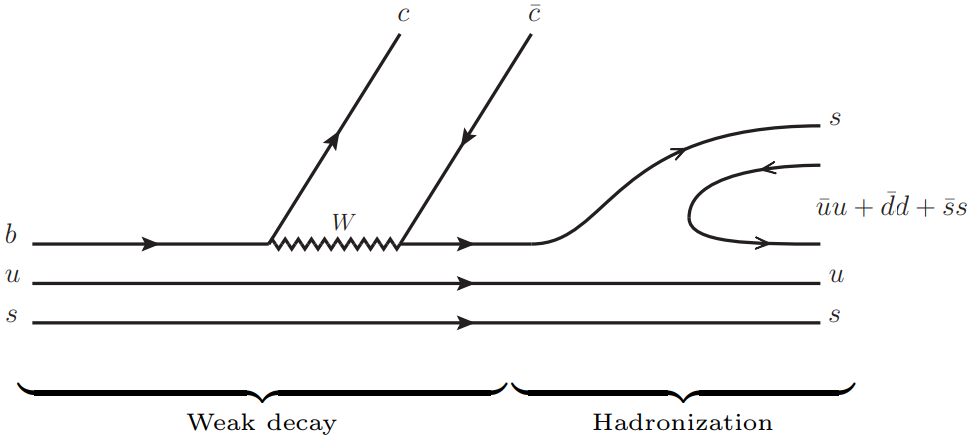}
    \caption{Microscopic quark level production of $J/\psi \ \phi \ \Lambda$ in the $\Lambda_b$ decay through internal conversion.}
    \label{int_con}
\end{figure}

The amplitude of the dominant process (Fig. \ref{Fig:X4160}) can be written as:
\begin{equation}
    {\cal M}^P_{X_{4160}}=A(\Vec{\epsilon}_{J/\psi}\times\Vec{\epsilon}_\phi)\cdot\Vec{P}_\Xi G_{D_s^*\Bar{D}_s^*}\frac{T_{D_s^*\Bar{D}_s^*,J/\psi \phi}}{g_{D_s^*\Bar{D}_s^*}g_{J/\psi \phi}} \ .
\label{eq:M4160}
\end{equation}
The constant $A$ on the right side of the equation represents the strength of the $\Xi_b\to\Xi D_s^*~\Bar{D}_s^*$ weak decay, shown in Fig. \ref{Fig:DD}. This value can be taken as a constant due to the limited range of energies involved in the $\Xi_b\to\Xi~J/\psi~\phi$ decay as it is argued in Ref. \cite{Feijoo:2015cca}. The $(\Vec{\epsilon}_{J/\psi}\times\Vec{\epsilon}_\phi)\cdot\Vec{P}_\Xi$ factor denotes the P-wave operator which is the minimum partial wave we need in the weak vertex to conserve the angular momentum, since the spins of the $\Xi_b$ and $\Xi$ are $J=1/2$ while that of the $X(4160)$ is $J=2$. Here $\vec{\epsilon}_i~(i={J/\psi}\,~\phi)$ represent the polarization of the vector mesons in the $J/\psi~\phi$ rest frame and $\Vec{P}_\Xi$ the tree-momentum of the $\Xi$ in the same frame. The factor $G_{D_s^*\Bar{D}_s^*}$ indicates the contribution of the $D_s^*\Bar{D}_s^*$ loop shown in Fig. \ref{Fig:X4160} and $T_{D_s^*\Bar{D}_s^*,J/\psi\phi}$ stands for the coupled-channel unitarized amplitude for the $D_s^*\Bar{D}_s^*\to J/\psi~\phi$ process. Note that we divide this amplitude by the couplings constants of the $X(4160)$ to the initial and final meson-baryon states. This is done for a proper comparison with the $X(4140)$ contribution, discussed in the following.

The alternative description of the decay process via the production of the $X(4140)$ is shown in Fig. \ref{Fig:X4140} and the associated amplitude reads:
\begin{equation}
    {\cal M}^P_{X_{4140}}=\frac{B}{2M_{X(4140)}[M_{J/\psi\phi}-M_{X(4140)}+i\frac{\Gamma_{X(4140)}}{2}]},\label{eq:M4140}
\end{equation}
where the $X(4140)$ resonance is parameterized with a Breit-Wigner shape, with a mass $M_{X(4140)}$ and a width $\Gamma_{X(4140)}$, and $M_{J/\psi\phi}$ is the  invariant mass of the $J/\psi\phi$ system. Note the appearance of a new constant $B$, connected to the strength of the $\Xi_b\to\Xi~X(4140)$ weak decay. As we want this constant to have the same units as $A$, we introduce a $2M_{X(4140)}$ factor in the denominator. In this process we do not need to have a P-vertex since the quantum number of the $X(4140)$ is $J=1$. Therefore the minimum partial wave needed to conserve angular momentum is $L=0$.

  \begin{figure}[h!]
     \includegraphics[scale=0.4]{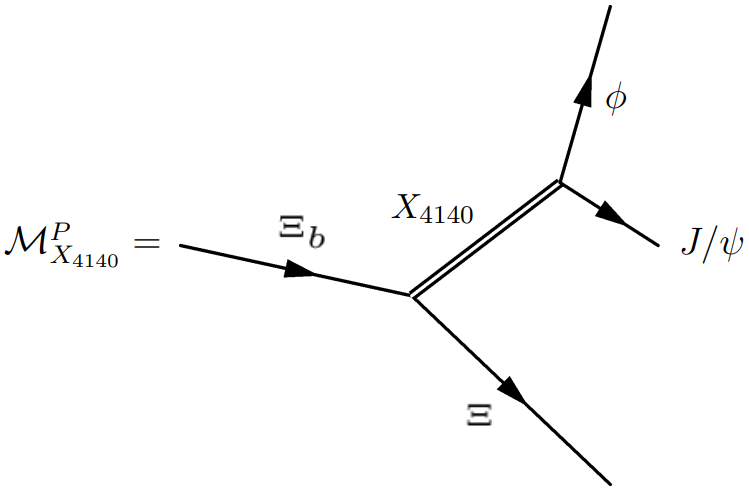}
     \caption{Mechanism for the $\Xi_b\to J/\psi~\phi~\Xi$ decay involving the $X(4140)$ resonance}
     \label{Fig:X4140}
 \end{figure}
 

\subsection{Final state interaction} \label{sec:FormFSI}
 
Once we discussed the generation of the $\Xi~J/\psi~\phi$ final state in the decay of the $\Xi_b$, we want to focus on studying the final state interaction of the $\Xi~J/\psi$ and $\Xi~\phi$ pairs.

  \begin{figure*}[t]
     \includegraphics[scale=0.35]{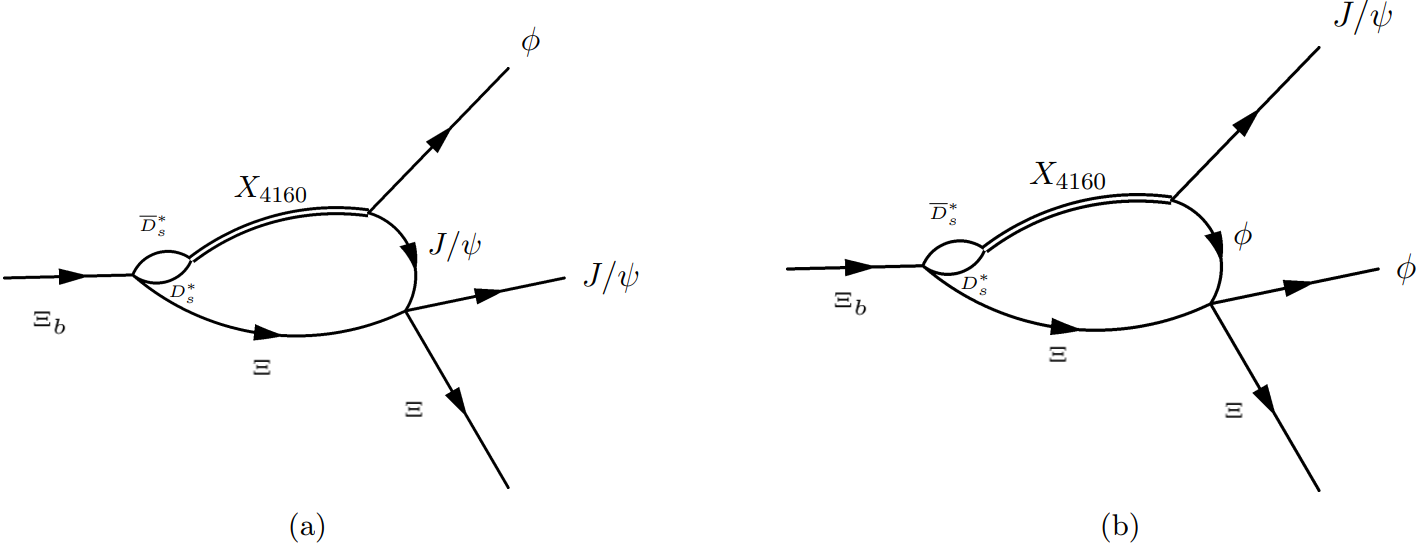}
     \caption{Final state interaction between $J/\psi~\Xi$ (a) and $\phi~\Xi$ (b) in the presence of the $X(4160)$ resonance.}
     \label{Fig:FSIX(4160)}
 \end{figure*}

The final state interaction of $\Xi~J/\psi$ pairs involves connecting the $\Xi$ and $J/\psi$ legs from the diagrams of Figs. \ref{Fig:X4160} and \ref{Fig:X4140}, forming a loop, which can dynamically generate the $S=-2$ pentaquark. 
The amplitude associated to the $\Xi_b$ decay process driven via the $X(4160)$ resonance including the final state interaction of the $J/\psi\Xi$ pair is:  
\begin{equation}
    {\cal M}^{J/\psi \Xi}_{X_{4160}}=A(\Vec{\epsilon}_{J/\psi}\times\Vec{\epsilon}_\phi)\cdot\bigg(\frac{\Vec{P}_\Xi-\Vec{P}_\phi}{2}\bigg)T_{J/\psi \Xi,J/\psi \Xi}I^{J/\psi \Xi}_{X_{4160}},
    \label{eq:M4160J/psi-Xi}
\end{equation}
which is associated to the diagram shown in Fig. \ref{Fig:FSIX(4160)}(a), where the superscript denotes the final state interaction channel. This amplitude is derived in Appendix \ref{Sec:App-A}, where the expression for the integral $I^{J/\psi \Xi}_{X_{4160}}$ is also given. We would like to remind the reader that for the decay via an intermediate $X(4160)$ to occur the weak vertex has to be in P-wave. Therefore the operator assigned to the the $\Xi_b\to D^*_s~\Bar{D}^*_s~\Xi$ is proportional to the momentum of the $\Xi$ and the loop integral (Fig. \ref{Fig:FSIX(4160)}(a)) becomes a tree-vector. We evaluate this loop integral in the $J/\psi~\Xi$ rest frame, where the loop integral becomes proportional to $\frac{(\Vec{P}_\Xi-\Vec{P}_\phi)}{2}$, with the scalar coefficient  $I^{J/\psi \Xi}_{X_{4160}}$. The term $T_{J/\psi \Xi,J/\psi \Xi}$, which captures the final state interaction contribution from the $J/\psi~\Xi$ pairs, is modeled using a Breit-Wigner form:
\begin{equation}
    T_{J/\psi \Xi,J/\psi \Xi}=\frac{g_{J/\psi~\Xi}^2}{M_{J/\psi~\Xi}-M+i\frac{\Gamma}{2}},
    \label{eq:T/psi-Xi}
\end{equation}
were $g_{J/\psi~\Xi}$ is the coupling of the $S=-2$ pentaquark to the $J/\psi~\Xi$ channel. $M_{J/\psi~\Xi}$ denotes the invariant mass of the $J/\psi~\Xi$ system, and $M$ and $\Gamma$ denote the mass and the width of the $S=-2$ pentaquark. The values of $g_{J/\psi~\Xi}$, $M$ and $\Gamma$ are taken from the molecular model of the VB interacting model discussed in Section \ref{sec:MBS}. 

We use Breit-Wigner formula, eq. (\ref{eq:T/psi-Xi}), instead of the full $T_{J/\psi \Xi,J/\psi \Xi}$ amplitude calculated in the previous section, because this form allows us to vary the position and the width of the pentaquark resonance in an easy and straightforward way, and, thus, to perform a more complete study on the possibility to observe double strangeness pentaquark in the $J/\psi~\Xi$ mass distribution from the $\Xi_b\to J/\psi~\phi~\Xi$ decay.

For the final state interaction of the $\phi~\Xi$ pair we can repeat similar calculations and obtain (see Appendix \ref{Sec:App-A}):
\begin{equation}
    {\cal M}^{\phi\Xi}_{X_{4160}}=A(\Vec{\epsilon}_{J/\psi}\times\Vec{\epsilon}_\phi)\cdot\bigg(\frac{\Vec{P}_\Xi+\Vec{P}_\phi}{2}\bigg)T_{\phi\Xi,\phi\Xi}I^{\phi\Xi}_{X_{4160}},
    \label{eq:M4160Phi-Xi}
\end{equation}
where the $T_{\phi\Xi,\phi\Xi}$ amplitude is also modeled with a Breit-Wigner form, as in Eq.~(\ref{eq:T/psi-Xi}), but in this case $M$ and $\Gamma$ are the experimental mass and width of the $\Xi(2500)$. Since we do not have any model which can give us the value of $g_{\phi\Xi}$, we will invert the problem and will study for which values of $g_{\phi\Xi}$ it is feasible to see this state using the $\Xi_b\to\Xi~J/\psi~\phi$ reaction.

Now we go to the final state interaction in the presence of the $X(4140)$ resonance. The corresponding diagrams are given in Figs. \ref{Fig:FSIX(4140)}(a) and \ref{Fig:FSIX(4140)}(b) and the associated analytical expressions are:
\begin{equation}
    {\cal M}^{J/\psi\Xi}_{X_{4140}}=\Tilde{B}T_{J/\psi\Xi,J/\psi\Xi}I^{J/\psi\Xi}_{X_{4140}},
    \label{eq:M4140J/Jpsi-Xi}
\end{equation}
\begin{equation}
    {\cal M}^{\phi\Xi}_{X_{4140}}=\Tilde{B}T_{\phi\Xi,\phi\Xi}I^{\phi\Xi}_{X_{4140}}.
    \label{eq:M4140Phi-Xi}
\end{equation}
The derivation of these expressions are explained in more detail in Appendix \ref{Sec:App-A}. The terms $I^{J/\psi\Xi}_{X_{4140}}$ and $I^{\phi\Xi}_{X_{4140}}$ are scalar loop integral analogous to those in Eqs. (\ref{eq:M4160J/psi-Xi}) and (\ref{eq:M4160Phi-Xi}).

  \begin{figure*}[t]
     \includegraphics[scale=0.35]{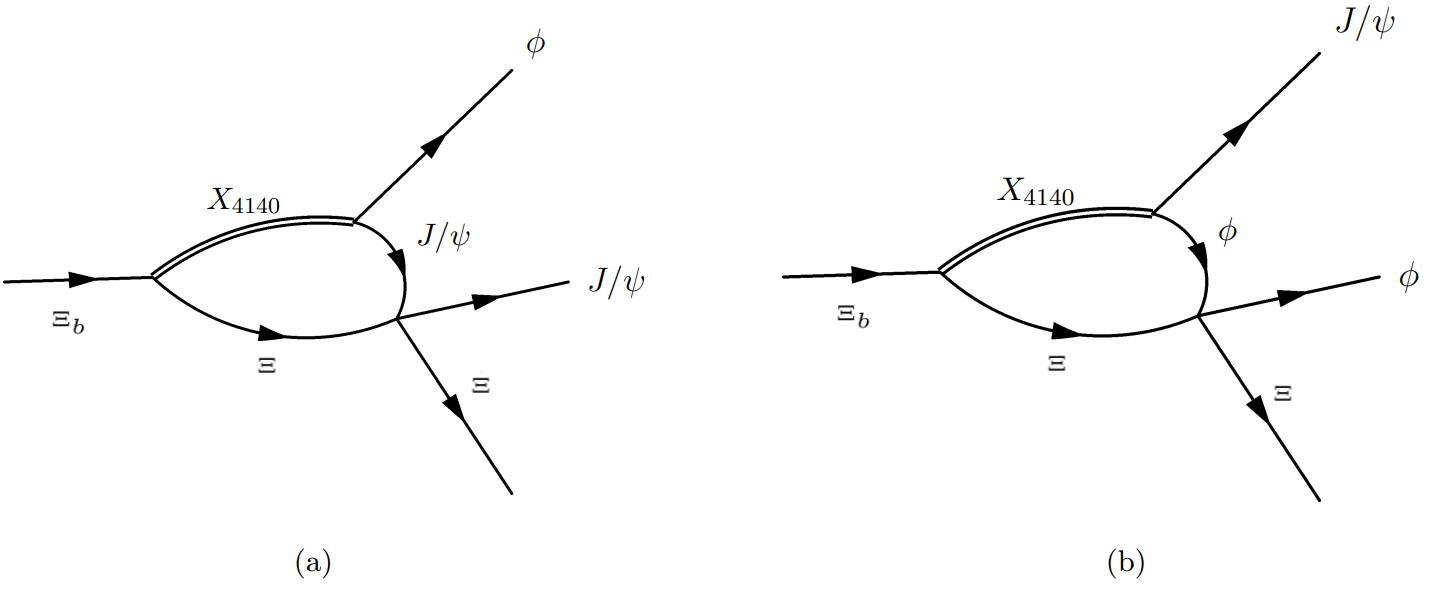}
     \caption{Final state interaction between $J/\psi~\Xi$ (a) and $\phi~\Xi$ (b) in the presence of the $X(4140)$ resonance.}
     \label{Fig:FSIX(4140)}
 \end{figure*}
 
 \subsection{Full amplitude and decay rate}

In this section we first proceed to combine the various terms to form the invariant amplitude in the cases where the decay driven by the $X(4140)$, ${\cal M}_{X_{4140}}$, or by the $X(4160)$,  ${\cal M}_{X_{41460}}$ :
 \begin{eqnarray}
    B\, {\cal M}_{X_{4140}}&=&{\cal M}^{P}_{X_{4140}}+{\cal M}^{J/\psi\Xi}_{X_{4140}}+{\cal M}^{\phi\Xi}_{X_{4140}}, 
    \label{eq:M4140tot} \\
    A\, {\cal M}_{X_{4160}}&=&{\cal M}^{P}_{X_{4160}}+{\cal M}^{J/\psi\Xi}_{X_{4160}}+{\cal M}^{\phi\Xi}_{X_{4160}},
    \label{eq:M4160tot}
\end{eqnarray}
where we wrote explicitly the primary decay strengths $A$ and $B$. 
Secondly, we add these terms obtaining the full amplitude denoted as ${\cal M}$,
\begin{eqnarray}
    \overline{|{\cal M}|^2}=|A|^2\overline{{|\cal M}_{4160}|^2}+|B|^2\overline{{|\cal M}_{4140}|^2} \nonumber \\ 
    =|A|^2\Big(\overline{{|\cal M}_{4160}|^2}+\beta\, \overline{{|\cal M}_{4140}|^2}\Big),
    \label{eq:Mbeta_0}
\end{eqnarray}
where the bar represents the sum over polarizations and $\beta=|B|^2/|A|^2$. Note that, since the weak decay goes in P-wave for the $X(4160)$ contribution and in S-wave for the $X(4140)$ one, the cross term in $\overline{{|\cal M}_{4160}+{\cal M}_{4140}|^2}$ cancels as these two partial waves are orthogonal and do not interfere. It is also important to comment that, although the the overall factor $|A|^2$ is not known, it is not relevant for the shape of the obtained distributions. The form of the distributions, and not its absolute value, will be our main observable. We are exploring if the position of the peak and its width can be measured in several model situations. So our final results will be given in arbitrary units and therefore we will set  $|A|^2=1$ from here to the end of this work:
\begin{equation}
    \overline{|{\cal M}|^2}=\overline{{|\cal M}_{4160}|^2}+\beta\, \overline{{|\cal M}_{4140}|^2}\,.
    \label{eq:Mbeta}
\end{equation}

The constant $\beta$ establishes a relative weight between the $X(4140)$ and $X(4160)$ contributions. Its value will be determined by following the procedure described in Ref. \cite{Wang:2017mrt}, where the interplay between the $X(4160)$ and the $X(4140)$ in the decay $B^+\to J/\psi~\phi~K^+$ is studied. The mechanism involving the $X(4160)$ in this reaction is rather similar at the quark level to our diagram shown in Fig. \ref{Fig:DD}, but without the spectator $s$ quark in the initial and final state. The topology of the diagrams is also similar in the case of $X(4140)$. Therefore the $\beta$ value in our reaction may be similar as that in Ref. \cite{Wang:2017mrt}, although some difference may arise since the partial waves involved in their weak decay vertex are P-wave and D-wave for the $X(4140)$ and $X(4160)$ resonances, respectively, compared to the S- and P-waves of the present work.

We next deal with the sum over the  polarizations represented by the bar in Eqs. (\ref{eq:Mbeta_0}) and (\ref{eq:Mbeta}). For the ${\cal M}_{X(4140)}$ amplitude this calculation is trivial and it only produces a constant, which can be reabsorbed into $\beta$. For the ${\cal M}_{X(4160)}$ amplitude the calculation is more complicated adding cross term contributions. 
First of all, we introduce the following definitions:
\begin{equation}
    {\cal M}^{P}_{4160}=(\Vec{\epsilon}_{J/\psi}\times\Vec{\epsilon}_\phi)\cdot\Vec{P}_\Xi~\Tilde{{\cal M}}^{P}_{4160},
    \label{eq:MPTilde}
\end{equation}
\begin{equation}
    {\cal M}^{J/\psi\Xi}_{4160}=(\Vec{\epsilon}_{J/\psi}\times\Vec{\epsilon}_\phi)\cdot\Vec{K}_2~\Tilde{{\cal M}}^{J/\psi\Xi}_{4160},
    \label{eq:MJpsiXiTilde}
\end{equation}
\begin{equation}
    {\cal M}^{\phi\Xi}_{4160}=(\Vec{\epsilon}_{J/\psi}\times\Vec{\epsilon}_\phi)\cdot\Vec{K}_1~\Tilde{{\cal M}}^{\phi\Xi}_{4160},
    \label{eq:MphiXiTilde}
\end{equation}
where $\Vec{K}_1\equiv(\Vec{P}_\Xi+\Vec{P}_\phi)/2$ and $\Vec{K}_2\equiv(\Vec{P}_\Xi-\Vec{P}_\phi)/2$. Then, as it is shown in Appendix \ref{Sec:App-B}, the sum over the polarizations leads to
\begin{eqnarray}
    \overline{|{\cal M}_{4160}|^2}=&|\Vec{P}_\Xi|^2|\Tilde{{\cal M}}^{P}_{4160}|^2+  \nonumber \\
    &|\Vec{K}_2|^2|\Tilde{{\cal M}}^{J/\psi\Xi}_{4160}|^2+|\Vec{K}_1|^2|\Tilde{{\cal M}}^{\phi\Xi}_{4160}|^2 \nonumber \\ 
    &+2\Vec{P}_\Xi\cdot\Vec{K}_2\Re(\Tilde{{\cal M}}^{P}_{4160}\Tilde{{\cal M}}^{*J/\psi\Xi}_{4160}) \\
    &+2\Vec{P}_\Xi\cdot\Vec{K}_1\Re(\Tilde{{\cal M}}^{P}_{4160}\Tilde{{\cal M}}^{*\phi\Xi}_{4160})\nonumber \\
    &+2\Vec{K}_1\cdot\Vec{K}_2\Re(\Tilde{{\cal M}}^{J/\psi}_{4160}\Tilde{{\cal M}}^{*\phi\Xi}_{4160}), \nonumber
    \label{eq:M4160simplify}
\end{eqnarray}
where $\Re$ is the real part of the complex argument.

Finally, the double differential cross-section for the  $\Xi_b \to  J/\Psi  \phi \Xi$ decay process reads \cite{rocamai}:
\begin{eqnarray}
& &\frac{d^2\Gamma}{dM_{\phi \Xi}dM_{J/\psi \Xi}}  =  \\
& &\frac{1}{{(2\pi)}^3}\frac{4M_{\Xi_b}M_{\Xi}}{32M_{\Xi_b}^3}\, \overline{|{\cal M}(M_{\phi \Xi},M_{J/\psi \Xi})}|^2\,
 2 M_{\phi \Xi}\,  2 M_{J/\psi \Xi}\, , \nonumber 
\label{eq:double_diff_cross}
\end{eqnarray}
where $\mathcal{M}$ is given by Eq.~(\ref{eq:Mbeta}). We wrote in Eq.~(\ref{eq:M4160simplify}) the double differential cross section as a function of the $M_{\phi \Xi}$ and $M_{J/\psi \Xi}$ invariant masses, but we could have used any other combination of the two particle invariant masses of the  $J/\Psi  \phi \Xi$ final state.  

Now, fixing the invariant mass $M_{J/\psi \Xi}$, one can integrate over $M_{\phi \Xi}$ in order to obtain $d\Gamma/dM_{J/\psi \Xi}$. In this case, the integration limits are given by:
\begin{eqnarray}
\left( M_{\phi \Xi}^2 \right)_{\rm max} &=&{\left(E^*_{\Xi}+E^*_{\phi}\right)}^2 \nonumber \\
&&-{\left(\sqrt{{E^*_{\Xi}}^2-M^2_{\Xi}}-\sqrt{{E^*_{\phi}}^2-m^2_{\phi}}\right)}^2
\end{eqnarray}
and
\begin{eqnarray}
\left( M_{\phi \Xi}^2 \right)_{\rm min}&=&{\left(E^*_{\Xi}+E^*_{\phi}\right)}^2 \nonumber \\
&&-{\left(\sqrt{{E^*_{\Xi}}^2-M^2_{\Xi}}+\sqrt{{E^*_{\phi}}^2-m^2_{\phi}}\right)}^2\,
\end{eqnarray}
where
\begin{equation}
E^*_{\Xi}=\frac{M_{J/\psi \Xi}^2-m_{J/\psi}^2+M^2_{\Xi}}{2M_{J/\psi \Xi}} \ ,
\end{equation}
\begin{equation}
E^*_{\phi}=\frac{M_{\Xi_b}^2-M_{J/\psi \Xi}^2-m^2_{\phi}}{2M_{J/\psi \Xi}} \ .
\end{equation}
Similar formulas are obtained when fixing the invariant mass $M_{\phi \Xi}$ and integrating over $M_{J/\psi \Xi}$ to obtain $d\Gamma/dM_{\phi \Xi}$.

\section{Exotic hadrons in $\Xi_b \to  J/\Psi  \phi \Xi$ decay}

\subsection{The $J/\psi~\phi$ mass distribution}

We start studying the $J/\psi~\phi$ mass distribution, which will allow us to investigate the $X(4140)/X(4160)$ interplay. We consider two possibles models: the first one, which is based in the latest experimental results \cite{LHCb:2016axx,LHCb:2016nsl}, assumes the existence of only one broad $X(4140)$ resonance with a pole position at $M_R=4146.5~{\rm MeV}$ and a width of $\Gamma=83~{\rm MeV}$, while the second model assumes the existence of a narrow $X(4140)$ plus a wide $X(4160)$ resonance, based on earlier experimental results \cite{Wu:2010vk,Hofmann:2005sw,Yuan:2012wz,Xiao:2013yca,Garcia-Recio:2013gaa} and a recent theoretical model \cite{Wang:2017mrt}.

We start with the model with a wide $X(4140)$. The corresponding results are shown in Fig. \ref{fig:J/psiPhi-X(4140)}, where we can see the $J/\psi~\phi$ invariant mass distribution for the tree level calculations (see the diagram of Fig. \ref{Fig:X4140}) and the corresponding spectrum when we also take into account the final-state interaction (FSI) diagrams of Fig. \ref{Fig:FSIX(4140)}. As we can see, the effect of FSI is rather small. This will be also true for the model with two $X$ resonances, so in this section we will further present only the results that include the FSI effects.

\begin{figure}[h!]
    \includegraphics[scale=0.38]{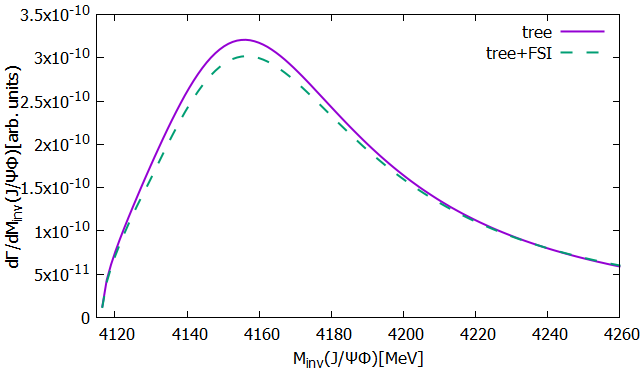}
    \caption{The $J/\psi~\phi$ mass spectrum computed using one wide $X(4140)$ resonance. The green line corresponds to tree level calculations only, while the purple one also includes the FSI effects.}
    \label{fig:J/psiPhi-X(4140)}
\end{figure}

We next discuss the results generated with two resonances, where, similarly to the results of Ref. \cite{Magas:2020zuo}, we adjust the parameter $\beta$ to a value which provides a similar strength for the contribution of the two $X$ resonances. The $X(4140)$ has a mass of $M_R=4132~{\rm MeV}$ and a width of $\Gamma=19~{\rm MeV}$, while the $X(4160)$ can be generated as a $D^*_s\bar{D}^*_s$ molecule, as we explained in Section \ref{sec:PrimDecay}. The corresponding model contains a loop-function that diverges and must be regularized. While a dimensional regularization scheme was employed in Ref. \cite{Molina:2009ct}, we proved earlier that such a scheme can produce fake poles. Therefore, in this work we use the cut-off regularization method with $\Lambda=650~{\rm MeV}$, which reproduces the results obtained using the dimensional regularization scheme in \cite{Molina:2009ct} in the energy region of interest. 

The $J/\psi~\phi$ invariant mass distribution for two $X$ model is displayed in Fig. \ref{fig:J/psiPhi-X(4140)X(4160)}, where we can see a narrow $X(4140)$ peak at $4132~{\rm MeV}$ and a rather wide $X(4160)$ peak. Note that the $X(4160)$ also produces a cusp around $4224~{\rm MeV}$, which corresponds to the  $D^*_s\bar{D}^*_s$ threshold. As it is mentioned in Ref. \cite{Magas:2020zuo,Wang:2017mrt}, if this cusp is experimentally detected it would strongly suggest a molecular interpretation of $X(4160)$, as well as the existence of a narrow $X(4140)$ and a wide $X(4160)$ resonance in this energy region.

\begin{figure}[h!]
    \includegraphics[scale=0.38]{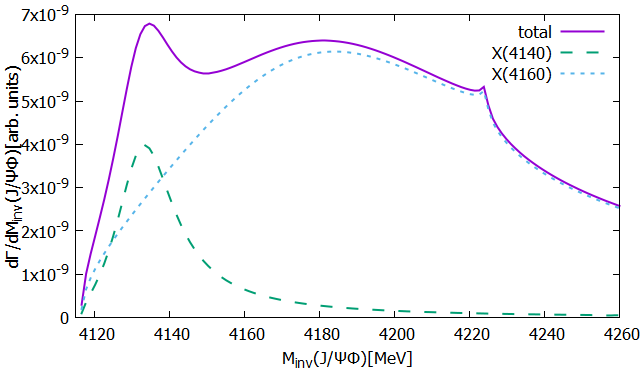}
    \caption{The purple line represents the $J/\psi~\phi$ mass spectrum generated using two resonance model. The green and the blue lines show the individual contributions from $X(4140)$ and $X(4160)$, respectively. }
    \label{fig:J/psiPhi-X(4140)X(4160)}
\end{figure}

For the results presented in Fig. \ref{fig:J/psiPhi-X(4140)X(4160)} the parameter $\beta$ of Eq.~(\ref{eq:Mbeta}) is taken as $\beta=\beta_0/2.6$, where $\beta_0$ is the value employed in Ref. \cite{Wang:2017mrt}. We would also like to analyze the sensibility of the obtained distribution with respect to the parameter $\beta$ and, for this purpose, we generated similar spectra for different values of $\beta$ and present these in Fig. \ref{fig:J/psiPhi-beta}. As it can be expected from Eq.~(\ref{eq:Mbeta}), changing the values of $\beta$ mainly affects the height of the $X(4140)$ peak.

\begin{figure}[h!]
    \includegraphics[scale=0.38]{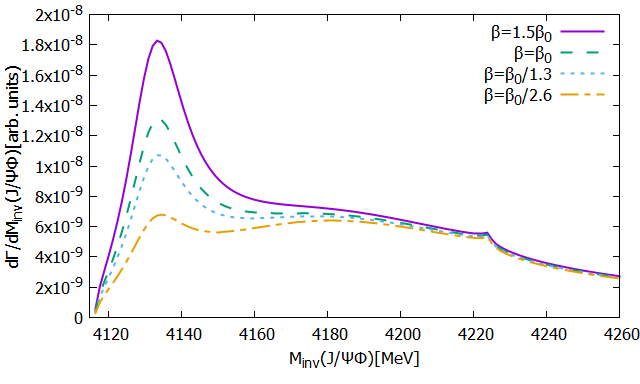}
    \caption{The solid lines represent the $J/\psi~\phi$ mass spectrum generated using the $X(4140)$ and the $X(4160)$ resonances for different values of the weight factor $\beta$, which controls the relative contributions of the two resonances; $\beta_0$ is the value of $\beta$ given in Ref. \cite{Wang:2017mrt}}.
    \label{fig:J/psiPhi-beta}
\end{figure}

\subsection{The $J/\psi~\Xi$ mass distribution}\label{sec:ResJpsiXi}

Now we discuss the results for the $J/\psi~\Xi$ mass distribution, which is the channel where we expect to see the signature of the $S=-2$ pentaquark. As seen in Table~\ref{tab:VBHigh}, the pentaquark generated dynamically from the VB interaction has a mass $M_R=4633~{\rm MeV}$, a width $\Gamma_R=80~{\rm MeV}$, and couples rather strongly to the $J/\psi~\Xi$ channel, with a value of $g_{J/\phi\Xi}=-1.62+0.38i$. We will take these values for the pole position and coupling as the nominal ones and we will vary them within a reasonable range to explore the sensitivity of our results to the pentaquark properties. We will also study the signal in the $J/\psi~\Xi$ distribution from a narrower pentaquark by using the parameters of the state obtained in Ref.~\cite{Roca:2024nsi}, namely a mass of $M_R=4617~{\rm MeV}$, a width of $\Gamma_R=12~{\rm MeV}$ and a coupling constant to the $J/\psi~\Xi$ channel of $g_{J/\phi\Xi}=0.66+0.0i$. 

\begin{figure}[h!]
    \includegraphics[scale=0.38]{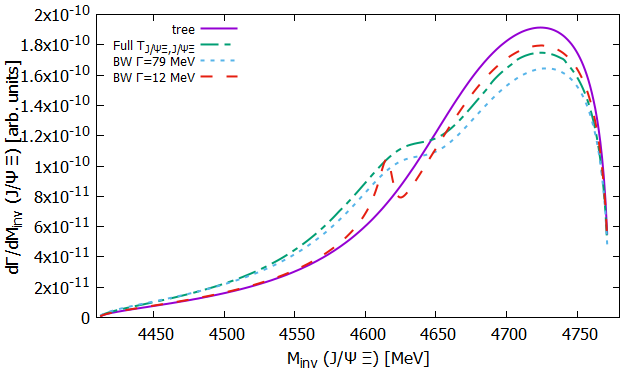}
    \caption{The $J/\psi~\Xi$ spectrum with a wide $X(4140)$ resonance: the purple line corresponds to the tree level diagram; 
    the green dash-dotted curve is generated by including the final state interaction effects using the full amplitude $T_{J/\phi\Xi,J/\psi\Xi}$ of the VB molecular model; 
    the blue dotted and the red dashed ones are obtained modeling the $S=-2$ pentaquark with a Breit-Wigner shape with $M_R=4633$~MeV and $\Gamma=80$~MeV or $M_R=4617$~MeV and $\Gamma=12$~MeV, respectively.}
    \label{fig:J/psiXi-4140}
\end{figure}

Proceeding in a similar manner as in the previous section, we start by showing our results computed with one wide $X(4140)$ resonance. In Fig. \ref{fig:J/psiXi-4140} we can see the $J/\psi~\Xi$ spectrum, where the solid line is obtained using only the tree-level diagram, while the other lines include also the FSI effects. In these latter curves we should see the signal of the $S=-2$ pentaquark. First of all, we note that the background shows itself as a wide structure around $4730~{\rm MeV}$. Both the green and blue lines present results for the molecular pentaquark with the caracteristics found by in our model, however the green dashed-dotted line is obtained using the full 
$T_{J/\psi \Xi,J/\psi \Xi}$ amplitude, while the blue dotted line is obtained by simulating this amplitude by 
a Breit-Wigner expression, Eq. (\ref{eq:T/psi-Xi}). The red dashed line is also a result of a simulation with the Breit-Wigner formula, however the parameters of the pentaquark state are taken from Ref.~\cite{Roca:2024nsi}. 

We observe that in our models the pentaquark interferes with the background positively and generates an important effect on the spectrum due to its sizable coupling to the $J/\psi~\Xi$ channel. The pentaquark obtained in the present work manifests itself as a bump around its nominal mass of $4633~{\rm MeV}$, not very prominent, but seemingly detectable.  The difference between green and blue curves comes from the fact that in the wide energy range the full amplitude may differ substantially from its Breit-Wigner approximation, and it gives the idea about the order of uncertainty which such an approximation may generate.   

The narrow pentaquark from \cite{Roca:2024nsi} interferes with the background negatively and leaves a sharper signal in the $J/\psi~\Xi$ invariant mass spectrum with approximately the same signal-to-noise ratio than the wider pentaquark.


\begin{figure}[h!]
    \includegraphics[scale=0.38]{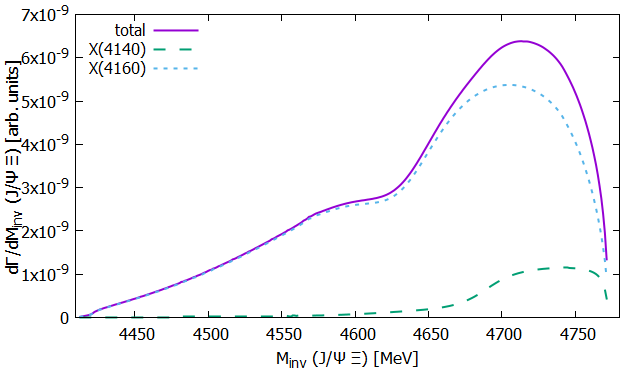}
    \caption{The purple line represents the $J/\psi~\Xi$ spectrum computed with a narrow $X(4140)$ plus a $X(4160)$, while the green and the blue ones show the individual contribution from the $X(4140)$ and the $X(4160),$ respectively.}
    \label{fig:J/psiXi-4140-4160}
\end{figure}

\begin{figure}[h!]
    \includegraphics[scale=0.38]{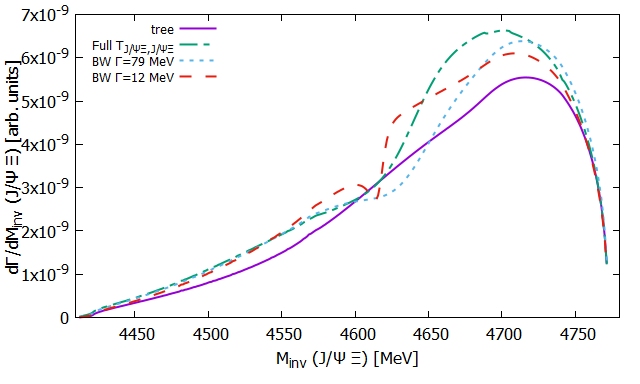}
    \caption{The $J/\psi~\Xi$ spectrum obtained with a narrow $X(4140)$ resonance plus the $X(4160)$: 
    the purple line corresponds to the tree level diagram; 
    the green dash-dotted curve is generated by including the final state interaction effects using the full amplitude $T_{J/\phi\Xi,J/\psi\Xi}$ of the VB molecular model; 
    the blue dotted and the red dashed ones are obtained modeling the $S=-2$ pentaquark with a Breit-Wigner shape with $M_R=4633$~MeV and $\Gamma=80$~MeV or $M_R=4617$~MeV and $\Gamma=12$~MeV, respectively.}
    \label{fig:J/psiXi-4140-4160_twoRes}
\end{figure}

We now consider the calculation with a narrow $X(4140)$ resonance and a wide $X(4160)$ one. The results including  FSI effects can be seen in Fig. \ref{fig:J/psiXi-4140-4160}, where the green and the blue curves show the individual contributions of the $X(4140)$ and the $X(4160)$ resonances, respectively, while the purple  solid curve is the sum of both. Note that the $X(4160)$ contribution is the dominant one. It is also important to see that, in contrast to the previous case, the signal of the pentaquark interferes negatively with the background. This can be seen more clearly in Fig. \ref{fig:J/psiXi-4140-4160_twoRes}, where we compare the spectra obtained assuming a wide or a narrow pentaquark. While the narrow option leaves an apparently better distinguishable signal, the presence of the $P_{css}$ is well detectable in both cases, with a similar signal-to-noise ratio.  Again the difference between the green and blue curves, actually more prominent that in case of a wide $X(4140)$ resonance, Fig. \ref{fig:J/psiXi-4140},  comes from the difference between employing the full $T_{J/\phi\Xi,J/\psi\Xi}$ amplitude and its Breit-Wigner approximation.

In the following results we will concentrate on the wider pentaquark, i.e. the one produced by our model and discussed in section \ref{sec:ResHeavy}.

In Fig. \ref{fig:J/psiXi-all}  we compare the the $J/\psi~\Xi$ spectrum for the one (left panels) and the two (right panels) $X$ resonance models including FSI, but varying the mass of the wide pentaquark and its coupling to the $J/\psi~\Xi$ channel, in order to implement the effect of some theoretical uncertainties. We modify the pentaquark pole position by $\pm50~{\rm MeV}$ (top to bottom panels) and we allow for a $20\%$ of variation of the coupling of the pentaquark to the channel $J/\psi~\Xi$ (different colors in each panel). The results of Fig.~\ref{fig:J/psiXi-all} clearly show that the presence of the double strangeness pentaquark is experimentally detectable in all cases. Actually, in the case of one wide $X(4140)$ (left panels) this task seems to be more easily achievable. And certainly the experimental measurement of the $J/\psi~\Xi$ invariant mass spectrum from the $\Xi_b\to\Xi~J/\psi~\phi$ decay would also allow one to distinguish, from the different shape of the pentaquark signal, between the one wide $X(4140)$ resonance model or that assuming two $X$ states.

\begin{figure*}[t]
    \includegraphics[scale=0.4]{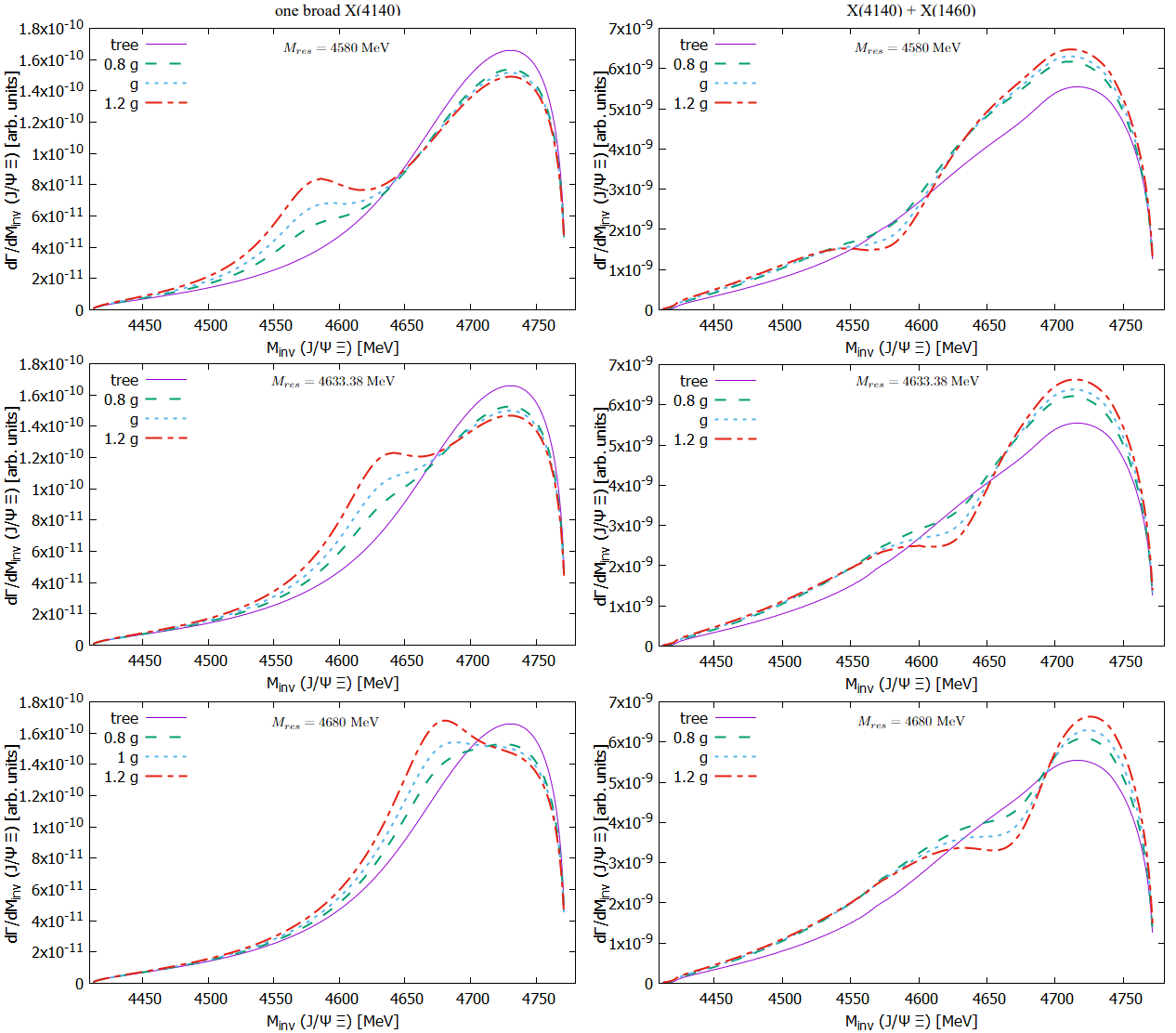}
    \caption{$J/\psi~\Xi$ spectrum computed with one broad $X(4140)$ resonance (left panels) or with a narrow $X(4140)$ plus a $X(4160)$  state (right panels) with different values of the $g_{J/\psi\Xi}$ coupling and different pole positions for the resonance [$4580~{\rm MeV}$ (top), $4663.38~{\rm MeV}$ (middle) and $4680~{\rm MeV}$ (bottom)]}
    \label{fig:J/psiXi-all}
\end{figure*}

\subsection{The $\phi~\Xi$ mass distribution} \label{sec:ResPhiXi}

Finally, we present the results for the $\phi~\Xi$ invariant mass distribution where we expect to see the signal of the $\Xi(2500)$. This appears as a one-star state in the PDG \cite{PDG}, with a mass of $M_{\Xi(2500)}=2500~{\rm MeV}$, and there exist two experimental groups that reported  different values for its width, namely $150~{\rm MeV}$ \cite{Alitti:1969rb} and $50~{\rm MeV}$ \cite{Aachen-Berlin-CERN-London-Vienna:1969bau}. Based on this information we decided to calculate the spectra shown in Fig. \ref{fig:phiXi-all}. They correspond to the different interpretation of the $X$ states: one wide $X(4140$) (left panels) or a narrow X(4140) plus a wide X(4160) (right panels). From top to bottom, we represent the spectrum for tree different widths, $\Gamma=50,~100,~150~{\rm MeV}$. Finally, as we do not know either which is the coupling constant of this state to the $\phi~\Xi$ channel, we explore a range of values for this coupling (different colors in each panel) to find the minimum value which would allow us to see the signal of the $\Xi(2500)$ in the generated spectrum.

\begin{figure*}[t]
    \includegraphics[scale=0.4]{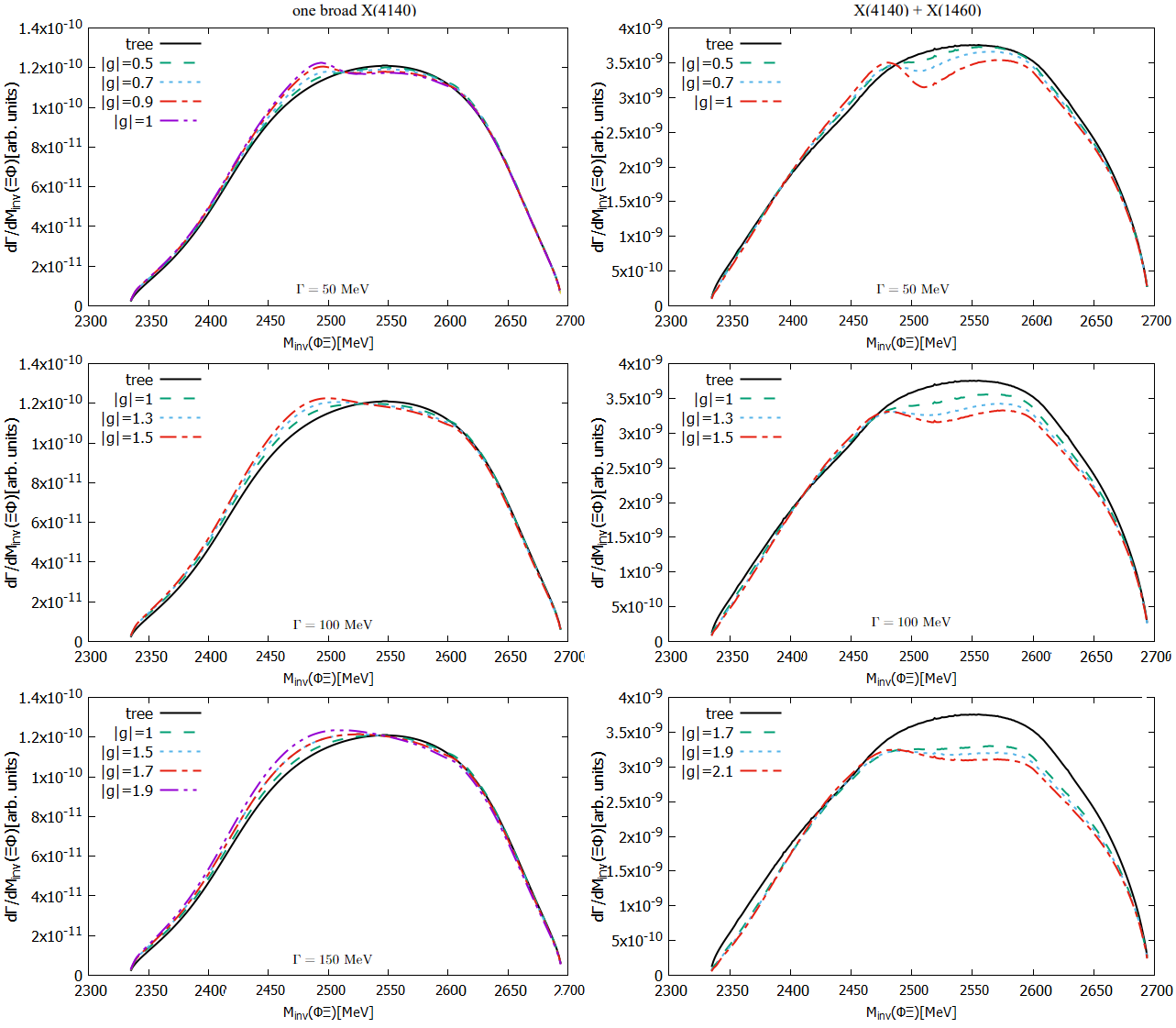}
    \caption{The $\phi~\Xi$ spectrum computed with one broad $X(4140)$ resonance (left panels) or with narrow $X(4140)$ plus a $X(4160)$ states (right panels) with different values for the coupling of the $\Xi(2500)$ to the $\phi~\Xi$ channel, and different widths of the resonance [$50~{\rm MeV}$ (top), $100~{\rm MeV}$ (middle) and $150~{\rm MeV}$ (bottom)]}
    \label{fig:phiXi-all}
\end{figure*}

We observe, as expected, that the minimum value of the coupling constant needed to generate a significant interference with the background is larger if the state has a larger width. More specifically, if $\Gamma=50~{\rm MeV}$ then the minimum absolute value of the coupling is around $0.7$, while for a $\Xi(2500)$ with $\Gamma=150~{\rm MeV}$ this value grows to $1.7-1.9$. Similarly to the previous case, the interference of the resonance with the background is positive for the case of one wide $X(4140)$ resonance, while for the case of a narrow $X(4140)$ resonance plus a wide $X(4160)$ one the interference is negative. Thus, studying this spectrum as a potential way of distinguishing between the two models would also be very interesting, provided the coupling of the $\Xi(2500)$ to the $\phi~\Xi$ states is large enough for its signal to be seen in the spectrum.

\section{Conclusions}

The discovery in the last twenty years of many exotic hadrons not fitting into the conventional quark composition model, has converted this topic into an important field of research. Motivated by this fact, in the present work we studied some resonances that could be interpreted as quasi-bound states of an interacting meson-baryon pair in the strangeness $-2$ and isospin $1/2$ sector, employing effective Lagrangians for describing the exchange of a vector meson in the t-channel. We also proposed a decay process $(\Xi_b\to\Xi~J/\psi~\phi)$ suitable to detect some of those exotic hadrons in all two-particle invariant mass spectra. 

First of all, in the light flavor sector, the scattering amplitude for the interaction of pseudoscalar mesons with baryons presents two poles, which can be associated to the $\Xi(1620)$ and $\Xi(1690)$ resonances from the PDG \cite{PDG} compilation. By modifying  some parameters in a reasonable range we can reproduce the experimental masses of the $\Xi(1620)$ and $\Xi(1690)$, producing two states at $1600~{\rm MeV}$ and $1685.40~{\rm MeV}$. We can therefore conclude that the $\Xi(1620)$ and the $\Xi(1690)$ may have a meson-baryon molecular origin. In our model, the $\Xi(1620)$ would have an important $\pi\Xi$ component, around $50\%$, with a $20\%$ mixture of $\Bar{K}\Lambda$. The $\Xi(1690)$ would be mainly a $\bar{K}\Sigma$ molecule with a small component $(2\%)$ of $\eta\Xi$. Note, however, that this composition differs from that of the states found by recent theoretical models that include the NLO terms of the Lagrangian, the widths of which are also more in accordance with the experimental values. This interesting topic deserves further investigations.

For the light-channel sector of the interaction of vector mesons with baryons we also find two poles in the scattering amplitude, which could be associated with two resonances listed in the PDG, namely the $\Xi(1820)$ and $\Xi(1950)$ \cite{PDG}. By modifying some parameters we can bring the positions of these two states closer to their experimental masses, but the theoretical widths are much smaller than the experimental ones. This might be solved by implementing the coupling between the vector meson channels with the pseudoscalar meson ones, thus opening the decay to lighter channels and correspondingly increasing the width of the dynamically generated states. If the present interpretation is assumed to represent the nature of these states, we could understand the $\Xi(1820)$ as a mixture of $\phi\Xi$ $(33\%)$ and $\bar{K}^*\Lambda$ $(13\%)$ components, while the $\Xi(1950)$ would be a mixture of $\Bar{K}^*\Sigma$ $(50\%)$, $\omega\Xi$ $(10\%)$ and $\phi\Xi$ $(10\%)$ ones. Note that these states are degenerate in spin-parity and form a $J^\pi=\frac{1}{2}^-$, $J^\pi=\frac{3}{2}^-$ doublet. This would be compatible with the observed quantum numbers of the $\Xi(1820)$, which are $J^\pi=\frac{3}{2}^-$.

Alternatively our model allows to generate these two resonances in the region of the $\Xi(1950)$, producing an apparent wide resonance. This is a  possibility pointed out in the PDG, where it is mentioned that the $\Xi(1950)$ may in fact be more than one $\Xi$ state.

The results for the heavy sector are quite encouraging, as they point towards the existence of double strangeness pentaquarks with hidden charm, which can be generated thanks to the coupled-channel structure of our model.  The absence in this sector
of a long range interaction mediated by pion-exchange
also makes the search for these states specially interesting. If they do exist, their interpretation as molecules
would require a change of paradigm, since they could
only be bound through heavier-meson exchange models
as the one employed in this work, 
 see Ref.~\cite{Marse-Valera:2022khy} for more details. 

Indeed, in the pseudoscalar meson baryon interaction sector we can generate one hidden charm $S=-2$ baryon with $M=4493~{\rm MeV}$ and $\Gamma=73~{\rm MeV}$, which can be interpret as a $\Bar{D}\Omega_c$ molecule, and the quantum numbers of this state would be $J^\pi=\frac{1}{2}^-$. In the vector meson baryon interaction sector we can also generate one state with $M=4633~{\rm MeV}$ and $\Gamma=80~{\rm MeV}$ and, similarly to the previous case, this state can be understood as a $\Bar{D}^*\Omega_c$ molecule which would be degenerate in spin-parity forming a $J^\pi=\frac{1}{2}^-$, $J^\pi=\frac{3}{2}^-$ doublet. 

The recent similar model of Ref. \cite{Roca:2024nsi} also finds such states, practically at the same mass, but with much narrower width, a fact tied to the smaller size of the coupled-channel coefficients. In a subsequent study, Ref.~\cite{Oset:2024fbk}, the authors suggest the $\Xi_b^0\to \eta \eta_c \Xi^0$ and $\Omega_b^-\to K^- \eta_c \Xi^0$ decay processes in order to observe the $P_{css}$ with $I(J^P)=\frac{1}{2}(\frac{1}{2}^-)$ generated via a pseudoscalar-baryon interaction. In contrast, in the present manuscript we have focused on the possibility to potentially detect a $P_{css}$ state generated from a vector-baryon interaction in the $\Xi_b\to\Xi~J/\psi~\phi$ decay process.

Another motivation to study the $\Xi_b\to\Xi~J/\psi~\phi$ decay process emerged from the inspection of the Dalitz plots of Fig. \ref{Fig:Dalitz}, which proved the potential possibility to study several exotic states that may be detected in all three two-body invariant mass spectra. Similarly to what is done in \cite{Magas:2020zuo}, we have developed two different models for the interpretation of the $X(4140)$ and $X(4160)$ resonances. The first model considers that there only exists one wide $X(4140)$ state while the other claims the existence of a narrow $X(4140)$ resonance plus a $X(4160)$ one. Consequently, in this latter case,  due to the nature of the $X(4160)$ as a $D^*_s\Bar{D}_s^*$ \cite{Molina:2009ct} bound system, there appears a cusp in the $J/\psi~\phi$ mass spectrum near the threshold of this channel $(4224~{\rm MeV})$. Therefore, if this cusp is detected, it would be a clear signal that the two-resonance model is the one that represents the nature of the $X$ resonances, and also that the $X(4160)$ is a molecular state.

The $X(4274)$, $X(4350)$ and the $X(4500)$ resonances may also leave a signal at higher $J/\psi$ invariant masses in the $\Xi_b\to\Xi~J/\psi~\phi$ decay, but they have not been the object of study in the present work.

In the $J/\psi~\Xi$ spectrum, we studied the possibility of detecting the double strange pentaquark with hidden charm, which we introduce as a Breit-Wigner using the parameters obtained from the coupled channel approach developed in this work. This permits us to keep the model simple and implement variations on the parameters easily. The results are very promising, since if the mass of this state lies in the region $4580-4680~{\rm MeV}$ it has a good chance to be experimentally detected in the spectrum of $J/\psi~\Xi$ pairs produced in the $\Xi_b\to\Xi~J/\psi~\phi$ decay. We have also shown that, in the case of a single wide $X(4140)$ model, the pentaquark signal interferes positively with the background, while, for the two $X$ model, this interference is negative. This potentially allows one to differentiate between these two situations once the corresponding spectrum will be measured. 

Finally, we also studied the $\phi~\Xi$ spectrum, where we explored the possibility to detect the $\Xi(2500)$. For this resonance there exists a discussion about whether its width is $50~{\rm MeV}$ or $150~{\rm MeV}$ \cite{PDG}, so we studied these two possibilities plus an intermediate case of $\Gamma=100~{\rm MeV}$. Our results show that, if the resonance has a narrow width of $50~{\rm MeV}$, its signal in the spectrum has high chances to be detected even with small coupling values of $g_{\phi\Xi}\sim0.7-0.9$, whereas if $\Gamma=100~{\rm MeV}$ the coupling should be larger than $g_{\phi\Xi}=1.3$, and even larger ($g_{\phi\Xi}\gtrsim 1.9$) if $\Gamma=150~{\rm MeV}$, in order to produce a significant signal above the background.

\section*{Acknowledgments}
This work is supported by the State Research Agency of the Ministry of Science, Innovation and Universities of Spain (MICIU/AEI/10.13039/ 501100011033) and by FEDER UE through
grant PID2023-147112NB-C21 and through the ``Unit of Excellence Mar\'ia de Maeztu 2020-2023" award to the Institute of Cosmos Sciences, grant CEX2019-000918-M. Additional support is provided by the Generalitat de Catalunya (AGAUR) through grant 2021SGR01095.

\appendix
\section{Appendix A: Double loop integral} \label{Sec:App-A}

In our discussion of the $\Xi_b\to\Xi~J/\psi~\phi$ decay we gave the expression for two final state interaction amplitudes ${\cal M}_{4160}^{J/\psi\Xi}$, and ${\cal M}_{4160}^{\phi\Xi}$ which involve a loop integral that is a 3-vector. In this appendix we demonstrate how we arrive to eqs. (\ref{eq:M4160J/psi-Xi}) and (\ref{eq:M4160Phi-Xi}). Similarly, we will derive the expressions for the ${\cal M}_{4140}^{J/\psi\Xi}$ and ${\cal M}_{4140}^{\phi\Xi}$ amplitudes.

\begin{figure}[h!]
     \includegraphics[scale=0.45]{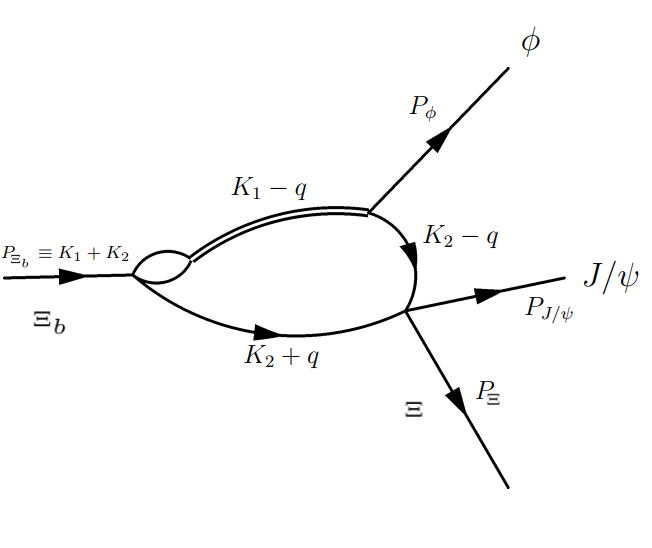}
     \caption{Feynman diagram associated to the ${\cal M}_{4160}^{J/\psi\Xi}$ amplitude, showing the momenta for all particle lines.}
     \label{Fig:fsi4160_mom}
\end{figure}
 
In Fig. \ref{Fig:fsi4160_mom} we can see the diagram associated to the amplitude ${\cal M}_{4160}^{J/\psi\Xi}$, where the momenta of each particle is shown. The corresponding expression is:
\begin{eqnarray}
     {\cal M}_{4160}^{J/\psi\Xi}=iT_{J/\psi\Xi,J/\psi\Xi}     \int\frac{d^4q}{(2\pi)^4}G_{D_s^*\Bar{D}_s^*} \nonumber\\
\frac{(\Vec{\epsilon}_{J/\psi}\times\Vec{\epsilon}_\phi)\cdot(\Vec{K}_2+\Vec{q})}{(K_1-q)^2-M_X^2+iM_X\Gamma_X} \nonumber\\
     \frac{1}{(K_2-q)^2-M_{J/\psi}^2+i\epsilon}  \nonumber\\ 
     \frac{2M_\Xi}{(K_2+q)^2-M_\Xi^2+i\epsilon},
\end{eqnarray}
where we have defined $K_1=(P_\Xi+P_\phi)/2$ and $K_2=(P_\Xi-P_\phi)/2$ (and correspondingly $K_1+K_2=P_\Xi$ and $K_1-K_2=P_\phi$). The $J/\psi\Xi$ scattering amplitude $T_{J/\psi\Xi,J/\psi\Xi}$ is introduced as a Breit-Wigner with the $S=-2$ pentaquark parameters. As discussed in section \ref{sec:FormFSI} the weak interaction has to be in P-wave what leads to the vector product of the corresponding polarization vectors times the momentum of the $\Xi$. Finally, the three internal lines correspond to the propagators of the $J/\psi$, $\Xi$ and $X(4160)$ resonances. Separating the $q^0$ integral and rearranging some terms we obtain:
\begin{eqnarray}
     {\cal M}_{4160}^{J/\psi\Xi}=iT_{J/\psi\Xi,J/\psi\Xi}(\Vec{\epsilon}_{J/\psi}\times\Vec{\epsilon}_\phi)\int\frac{d^3q}{(2\pi)^3}(\Vec{K}_2+\Vec{q}) \nonumber \\
     \int\frac{dq^0}{2\pi}\frac{G_{D_s^*\Bar{D}_s^*}}{(K_1^0-q^0)^2-\omega_X^2+iM_X\Gamma_X} \nonumber \\
     \frac{1}{(K_2^0-q^0)^2-\omega_{J/\psi}^2+i\epsilon} \frac{2M_\Xi}{(K_2^0+q^0)^2-\omega_\Xi^2+i\epsilon}, \nonumber \\
\end{eqnarray}
where $\omega_X^2=M_X^2+(\Vec{K}_1-\Vec{q})^2$, $\omega_{J/\psi}^2=M_{J/\psi}^2+(\Vec{K}_2-\Vec{q})^2$ and $\omega_{\Xi}^2=M_\Xi^2+(\Vec{K}_2+\Vec{q})^2$. Taking a non relativistic limit, this expression can be written as:
\begin{eqnarray}
     {\cal M}_{4160}^{J/\psi\Xi}=  iT_{J/\psi\Xi,J/\psi\Xi}(\Vec{\epsilon}_{J/\psi}\times\Vec{\epsilon}_\phi)\int\frac{d^3q}{(2\pi)^3}(\Vec{K}_2+\Vec{q}) \nonumber \\
     \int\frac{dq^0}{2\pi}\frac{G_{D_s^*\Bar{D}_s^*}}{2\omega_X[K_1^0-q^0-\omega_X+i\frac{\Gamma_X}{2}]} \nonumber \\
      \frac{1}{2\omega_{J/\psi}[K_2^0-q^0-\omega_{J/\psi}+i\epsilon]}\frac{M_\Xi}{\omega_\Xi[K_2^0+q^0-\omega_\Xi+i\epsilon]} \ , \nonumber \\
\end{eqnarray}
where the poles of the integrand  can be easily located:
\begin{eqnarray}
     q^0=K_1^0-\omega_X+i\frac{\Gamma}{2} & \rightarrow & {\rm Resonance}, \\ 
     q^0=K_2^0-\omega_{J/\psi}+i\epsilon & \rightarrow &  J/\psi~{\rm propagator}, \\ 
     q^0=-K_2^0+\omega_\Xi-i\epsilon & \rightarrow &  \Xi~{\rm propagator}.
\end{eqnarray}

Since there are two poles above the real axis and one below, we compute the integral closing the contour in the lower half plane, thus picking only the $\Xi$ propagator pole and finding:
\begin{eqnarray}
     {\cal M}_{4160}^{J/\psi\Xi}=T_{J/\psi\Xi,J/\psi\Xi}(\Vec{\epsilon}_{J/\psi}\times\Vec{\epsilon}_\phi) \nonumber \\
     \int\frac{d^3q}{(2\pi)^3}\frac{(\Vec{K}_2+\Vec{q})G_{D_s^*\Bar{D}_s^*}}{2\omega_X[K_1^0+K_2^0-\omega_\Xi-\omega_X+i\frac{\Gamma_X}{2}]} \nonumber \\ 
     \frac{1}{2\omega_{J/\psi}[2K_2^0-\omega_\Xi-\omega_{J/\psi}+i\epsilon]}\frac{M_\Xi}{\omega_\Xi}.
\end{eqnarray}

To compute this integral we work in the Jackson frame (the $J/\psi$ $\phi$ rest frame), where $\Vec{K}_1$ and $\Vec{K}_2$ are almost equal since $P_\phi$ is small. So we can can replace $\Vec{K}_1$ with $\Vec{K}_2$, and thus now the integral only depends on $\Vec{K}_2$. In this way the integral can be taken to be proportional to this vector, giving:
\begin{equation}
     {\cal M}_{4160}^{J/\psi\Xi}=T_{J/\psi\Xi,J/\psi\Xi}(\Vec{\epsilon}_{J/\psi}\times\Vec{\epsilon}_\phi)\Vec{K}_2I^{J/\psi\Xi}_{X(4160)},
\end{equation}
with,
\begin{eqnarray}
    I^{J/\psi\Xi}_{X(4160)}=  \int\frac{d^3q}{(2\pi)^3}\frac{\Vec{K}_2\cdot\Vec{q}~G_{D_s^*\Bar{D}_s^*}}{2\omega_X[K_1^0+K_2^0-\omega_\Xi-\omega_X+i\frac{\Gamma_X}{2}]} \nonumber \\
      \frac{1}{|\Vec{K}_2|^2}
    \frac{1}{2\omega_{J/\psi}[2K_2^0-\omega_\Xi-\omega_{J/\psi}+i\epsilon]}\frac{M_\Xi}{\omega_\Xi}, \nonumber \\
\end{eqnarray}
where the shift $\Vec{q}\to\Vec{q}+\Vec{K}_2$ has been made, and $\omega_X=M_X^2+(2\Vec{K}_2-\Vec{q})^2$, $\omega_{J/\psi}=M_{J/\psi}^2+(2\Vec{K}_2-\Vec{q})^2$ and $\omega_{\Xi}=M_\Xi^2+\Vec{q}^2$.

Once we have ${\cal M}_{4160}^{J/\psi\Xi}$, the evaluation of ${\cal M}_{4160}^{\phi\Xi}$ is rather straightforward, as one only needs to change $J/\psi$ to $\phi$ and vice versa. In the Jackson frame this is equal to the interchange of $\Vec{K}_2$ and $\Vec{K}_1$, leading to:
\begin{equation}
     {\cal M}_{4160}^{\phi\Xi}=T_{\phi\Xi,\phi\Xi}(\Vec{\epsilon}_{J/\psi}\times\Vec{\epsilon}_\phi)\Vec{K}_1I^{\phi\Xi}_{X(4160)},
\end{equation}
with,
\begin{eqnarray}
    I^{\phi\Xi}_{X(4160)}=\int\frac{d^3q}{(2\pi)^3}\frac{\Vec{K}_1\cdot\Vec{q}~G_{D_s^*\Bar{D}_s^*}}{2\omega_X[K_1^0+K_2^0-\omega_\Xi-\omega_X+i\frac{\Gamma_X}{2}]}\nonumber \\ 
     \frac{1}{|\Vec{K}_1|^2}\frac{1}{2\omega_{\phi}[2K_1^0-\omega_\Xi-\omega_{\phi}+i\epsilon]}\frac{M_\Xi}{\omega_\Xi}, \nonumber \\
\end{eqnarray}
where $\omega_X=M_X^2+(2\Vec{K}_1-\Vec{q})^2$, $\omega_{\phi}=M_{\phi}^2+(2\Vec{K}_1-\Vec{q})^2$ and $\omega_{\Xi}=M_\Xi^2+\Vec{q}^2$.

Finally, we can also derive the expressions for ${\cal M}_{4140}^{J/\psi\Xi}$ and ${\cal M}_{4140}^{\phi\Xi}$ proceeding in an analogous way, with the difference that now we do not have the $G_{D^*_s\Bar{D}^*_s}$ loop-function and all interactions are in S-wave, and consequently these amplitudes are scalars. Hence, we can write:
\begin{equation}
     {\cal M}_{4140}^{J/\psi\Xi}=T_{J/\psi\Xi,J/\psi\Xi}I^{J/\psi\Xi}_{X(4140)},
\end{equation}
with,
\begin{eqnarray}
    I^{J/\psi\Xi}_{X(4140)}= \int\frac{d^3q}{(2\pi)^3}\frac{1}{2\omega_X[K_1^0+K_2^0-\omega_\Xi-\omega_X+i\frac{\Gamma_X}{2}]} \nonumber \\
    \frac{1}{2\omega_{J/\psi}[2K_2^0-\omega_\Xi-\omega_{J/\psi}+i\epsilon]}\frac{M_\Xi}{\omega_\Xi}, \nonumber \\
\end{eqnarray}
where $\omega_X=M_X^2+(2\Vec{K}_2-\Vec{q})^2$, $\omega_{J/\psi}=M_{J/\psi}^2+(2\Vec{K}_2-\Vec{q})^2$ and $\omega_{\Xi}=M_\Xi^2+\Vec{q}^2$, and
\begin{equation}
     {\cal M}_{4140}^{\phi\Xi}=T_{\phi\Xi,\phi\Xi}I^{\phi\Xi}_{X(4140)},
\end{equation}
with,
\begin{eqnarray}
    I^{\phi\Xi}_{X(4140)}= \int\frac{d^3q}{(2\pi)^3}\frac{1}{2\omega_X[K_1^0+K_2^0-\omega_\Xi-\omega_X+i\frac{\Gamma_X}{2}]}\nonumber \\ 
     \frac{1}{2\omega_{\phi}[2K_1^0-\omega_\Xi-\omega_{\phi}+i\epsilon]}\frac{M_\Xi}{\omega_\Xi}, \nonumber \\
\end{eqnarray}
where $\omega_X=M_X^2+(2\Vec{K}_1-\Vec{q})^2$, $\omega_{\phi}=M_{\phi}^2+(2\Vec{K}_1-\Vec{q})^2$ and $\omega_{\Xi}=M_\Xi^2+\Vec{q}^2$.

\section{Appendix B: Spin Sums} \label{Sec:App-B}

In order to calculate the final mass distribution of the $\Xi_b\to\Xi~J/\psi~\phi$ decay we needed to compute the average $\overline{|{\cal M}_{4160}}|^2$. In this appendix we show in more detail how we arrived to our final result, Eq.~(\ref{eq:M4160simplify}).  We start with the ${\cal M}_{4160}$ amplitude in the following form:
\begin{equation}
    {\cal M}_{4160}=(\Vec{\epsilon}_{J/\psi}\times\Vec{\epsilon}_\phi)\cdot(\Vec{P}_\Xi\Tilde{{\cal M}}_{4160}^P+\Vec{K}_2\Tilde{{\cal M}}_{4160}^{J/\psi}+\Vec{K}_1\Tilde{{\cal M}}_{4160}^{\phi}),
\end{equation}
where $\Vec{K}_1=(\Vec{P}_\Xi+\Vec{P}_\phi)/2$ and $\Vec{K}_2=(\Vec{P}_\Xi-\Vec{P}_\phi)/2$. Now we take the square of the absolute value of ${\cal M}_{4160}$,
\begin{eqnarray}\label{eq:M4160index}
    |{\cal M}_{4160}|^2&=&\varepsilon^{ijk}\epsilon_{J/\psi}^j\epsilon_\phi^k\varepsilon^{abc}\epsilon_{J/\psi}^b\epsilon_\phi^c  \\
    &\times&(P_\Xi^i\Tilde{{\cal M}}_{4160}^P+K_2^i\Tilde{{\cal M}}_{4160}^{J/\psi}+K_1^i\Tilde{{\cal M}}_{4160}^{\phi})\nonumber \\
    &\times& (P_\Xi^a\Tilde{{\cal M}}_{4160}^{*P}+K_2^a\Tilde{{\cal M}}_{4160}^{*J/\psi}+K_1^a\Tilde{{\cal M}}_{4160}^{*\phi}), \nonumber
\end{eqnarray}
where  $\varepsilon^{ijk}$ is the Levi-Civita symbol.

Note that, since we are working in a reference frame where the $J/\psi~\phi$ system is at rest, the 3-momenta of all these particles are small in comparison with their masses, therefore we can take a non-relativistic limit, where 
\begin{equation}
 \sum_{pol}\epsilon_{J/\psi}^j\epsilon_{J/\psi}^b=\delta^{jb},
\quad  \sum_{pol}\epsilon_{\phi}^k\epsilon_{\phi}^c=\delta^{kc}.
\end{equation}

Hence, performing the sum over the polarizations, Eq.~(\ref{eq:M4160index}) can be simplified to:
\begin{eqnarray}
    \overline{|{\cal M}_{4160}|^2}&=&\varepsilon^{ijk}\varepsilon^{ajk}(P_\Xi^i\Tilde{{\cal M}}_{4160}^P+K_2^i\Tilde{{\cal M}}_{4160}^{J/\psi}+K_1^i\Tilde{{\cal M}}_{4160}^{\phi}) \nonumber \\
  & \times & (P_\Xi^a\Tilde{{\cal M}}_{4160}^{*P}+K_2^a\Tilde{{\cal M}}_{4160}^{*J/\psi}+K_1^a\Tilde{{\cal M}}_{4160}^{*\phi}). 
\end{eqnarray}

Now, applying the $\varepsilon^{ijk}\varepsilon^{ajk}=\delta^{ia}$ property, we obtain
\begin{eqnarray}
    \overline{|{\cal M}_{4160}|^2}=(\Vec{P}_\Xi\Tilde{{\cal M}}_{4160}^P+\Vec{K}_2\Tilde{{\cal M}}_{4160}^{J/\psi}+\Vec{K}_1\Tilde{{\cal M}}_{4160}^{\phi}) \nonumber \\
  \times  (\Vec{P}_\Xi\Tilde{{\cal M}}_{4160}^{*P}+\Vec{K}_2\Tilde{{\cal M}}_{4160}^{*J/\psi}+\Vec{K}_1\Tilde{{\cal M}}_{4160}^{*\phi}). 
\end{eqnarray}
From this expression we can straightforwardly derive Eq.~(\ref{eq:MphiXiTilde}) by just computing the scalar products.

\end{document}